%% file: main.tex
\documentclass[sigplan,10pt,nonacm]{acmart}
\input{preamble-common.tex}

\begin{document}

\title{The Output-Space Hypothesis: Enumerative Equivalence Checking for Tensor Programs}

\author{Paul Biberstein}
\email{paulbib@seas.upenn.edu}
\orcid{0009-0001-5637-5878}
\affiliation{%
  \institution{University of Pennsylvania}
  \city{Philadelphia}
  \state{PA}
  \country{USA}
}
\author{Joseph Devietti}
\email{devietti@seas.upenn.edu}
\orcid{0000-0002-9330-7233}
\affiliation{%
  \institution{University of Pennsylvania}
  \city{Philadelphia}
  \state{PA}
  \country{USA}
}
\author{Mayur Naik}
\email{mhnaik@seas.upenn.edu}
\orcid{0000-0003-1348-8618}
\affiliation{%
  \institution{University of Pennsylvania}
  \city{Philadelphia}
  \state{PA}
  \country{USA}
}

\begin{abstract}

Tensor programs, as used in deep learning models, are a prime target for optimization, as small performance improvements can have a large impact across training or inference workloads.
However, such optimizations are complicated and can produce subtle bugs.
Traditionally, correctness is assumed when differential testing against a reference on random inputs fails to reveal bugs.
However, the inputs to these programs are massive tensors, and finding bugs can require generating extremely low likelihood inputs with precise relationships among their values.

We propose a novel way to find bugs more consistently by flipping the quantifiers.
Rather than generating a single input and checking all output tensor locations for equivalence, what if you could check a single output tensor location's equivalence for all inputs?
We implement this idea in a system, \dirigo, by using a novel symbolic execution strategy.
We demonstrate that \dirigo can find bugs effectively in a public dataset of 6,988 AI-written CUDA kernels that are all marked correct by differential testing. Of these, \dirigo finds 600 kernels that are actually buggy, and finds 97.3\% of those bugs within two minutes.

\end{abstract}

\maketitle

\section{Introduction}
\label{sec:intro}

Deep learning models are generally written in tensor programming frameworks like PyTorch, which offer the flexibility of a high-level language while maintaining workable performance.
If more performance is needed, these frameworks allow parts of the program to be replaced with highly-optimized GPU kernels.
To avoid bugs when writing low-level GPU kernels, we need automated systems that can perform end-to-end equivalence-checking on tensor programs even in the presence of these custom kernels. While many works port traditional CPU testing methods to the GPU \cite{gklee2012,gpuverify2012,collingbourne2012kleecl2}, they take a per-kernel view and do not scale or support heterogeneous programs that mix high-level PyTorch operations with custom kernels.

In the absence of suitable equivalence-checking, the standard benchmark for tensor program optimization, KernelBench~\cite{kernelbench2025}, adopts a simple correctness check based on \emph{differential testing}---comparing the optimized kernel's output against a reference implementation's up to a hand-tuned tolerance. As the left side of \autoref{fig:intro} illustrates, differential testing checks few of the many possible inputs, so it can miss bugs triggered by specific inputs and deem a buggy kernel correct. Moreover, KernelBench is also the primary benchmark used by the burgeoning field of agentic kernel optimization~\cite{autocomp2025,sakana2025,kevin2025,cudal1_2025,kernelevolve2025,cutegen2026}, and agents frequently ``reward-hack''~\cite{sakana_robust2025,metr2025} and produce fast but incorrect kernels that slip past differential testing. Formal verification is a natural antidote, but formal techniques struggle to scale to real tensor programs, even though these are generally specialized to fixed input and output shapes and lack data-dependent control-flow or indirect memory accesses.

\input{fig-intro}
\input{fig-example}

Our key insight is that, while bugs are sparse in the input space, they are often dense in the \emph{output space} of a tensor program---a buggy input will trigger incorrect values at many output locations, not just one. We call this the \textbf{Output-Space Hypothesis}: bugs are much denser in the output space of a tensor program than in input space. But how can we leverage this insight to find bugs more efficiently?

Our system, \dirigo, uses a scalable symbolic execution framework to enumerate the dataflow within a mixed PyTorch/CUDA tensor program, allowing us to work backwards from one output location to all the input locations that can affect it (right side of \autoref{fig:intro}). \dirigo assembles an equivalence condition and uses SMT solving to check whether an optimized kernel and a reference implementation produce the same values, at a given output location, for all possible values of the upstream inputs. Checking a handful of output locations offers a huge tractability advantage when output tensors have millions of elements. Thanks to \hypothesis, sparse checking suffices to find many bugs. Checking can also be parallelized across output locations, offering a path to full equivalence-checking with sufficient compute.

We use \dirigo to search for bugs in the 6,988 agent-generated KernelBench level-2 solutions in the \archive~\cite{sakana2025} that KernelBench's differential testing marks correct. We find that 600 (8.6\%) are in fact buggy, revealing the brittleness of differential testing in practice.

This paper makes the following contributions:
\begin{enumerate}
  \item \Hypothesis: the bugs likely to arise in tensor programs are sparse in the input-value space but dense in the output-location space.
  \item \dirigo, a scalable method for symbolic execution and equivalence-checking that makes per-location equivalence checks feasible for tensor programs.
  \item 600 new bugs in the \archive's solutions to the KernelBench level-2 problems, alongside 1,646 solutions shown to be fully correct across all output locations.
  \item Experimental evidence that \hypothesis holds for our dataset.
\end{enumerate}

\section{Illustrative Overview}
\label{sec:overview}

We now walk through an illustrative example of tensor program optimization via custom kernels to show how our hypothesis plays out on a real optimization trajectory.
Consider the reference program in \autoref{fig:example}, which is task 36 from the KernelBench level-2 benchmark.
The program is representative of typical deep learning workloads: it takes an input tensor \texttt{x} and weights and biases \texttt{W} and \texttt{b} and computes their convolution, then a channel-wise minimum and a height-wise sum, and finally applies \GELU (an activation function)~\cite{hendrycks2016gelu} and adds the bias.

This program is a good example of the restricted subset of programs we target: control flow, memory access addresses, and shapes of all intermediate tensors are data independent.
In other words, the order of operations executed and shapes of their results are constant under any input data, so long as the input shapes are unchanged.
This is a common pattern in deep learning workloads:
\begin{enumerate}
  \item KernelBench, the dominant benchmark in kernel optimization, exhibits this property for all level-1, level-2, and level-3 tasks (level-4 has only a specification, no stable implementation).
  \item PyTorch and JAX, the two most popular deep learning frameworks, only accelerate a computation (with \texttt{torch.compile} and \texttt{jax.jit} respectively) if it has this property, indicating the prevalence of code that matches this restriction.
\end{enumerate}

\input{fig-overview}

Now compare the reference program with the optimized version of the same computation taken from the AI CUDA Engineer Dataset.
Here, the optimized implementation calls the same PyTorch function in three out of five cases, but fuses the \texttt{min} and \texttt{sum} calls into an optimized kernel in CUDA to avoid a round-trip to global memory.
The kernel showcases features unique to GPU programming that pose challenges for analysis:
\begin{enumerate}
  \item \textbf{Large Thread Grid.} The kernel is launched with a grid of $128\times64$ blocks, each with 256 threads, for a total of 2,097,152 threads. This is common in GPU programming, but makes it infeasible to analyze each thread individually.
  \item \textbf{Shared Memory.} The kernel uses shared memory (a programmer-managed scratchpad backed by the same memory as the L1 cache) to hold each thread's partial sum of channel-wise minimums while they are reduced across the block. This is a common optimization, and is challenging to analyze since shared memory is shared across threads in a block, but not between blocks.
  \item \textbf{Warp-Level Reductions.} The kernel finishes the sum with a warp-level reduction across 32 threads (a warp). Again, cross-thread communication poses a challenge.
\end{enumerate}
As a result the optimized kernel is not equivalent to the reference implementation: the first loop that accumulates the sum in shared memory has the following configuration:
\begin{lstlisting}[style=inline,language=C++]
for (unsigned s = nT/2; s > 32; s >>= 1)
\end{lstlisting}
But it should have the adjusted stop condition:
\begin{lstlisting}[style=inline,language=C++]
for (unsigned s = nT/2; s >= 32; s >>= 1)
\end{lstlisting}
Without this fix, the loop misses the last step of accumulating 64 partial results into 32 partial results, so the final intra-warp summation only reads half-sums from shared memory, resulting in a final sum that is incorrect.

This bug is simple, but proves very challenging for differential testing, since exposing it requires complicated relationships among the tensor's elements. Informally, running the reference PyTorch with \kernelbench's default input distribution produces a post-convolution tensor whose values are roughly Gaussian with mean 0 and standard deviation 0.13. The channel-wise \texttt{min} over 16 channels produces values that are largely in the range $[-0.4,-0.1]$, and the \texttt{sum}, even operating on only half the column, produces a negative value in the range $[-10,-4]$. For the difference to clear \kernelbench's $10^{-2}$ tolerance, the half-column sum must exceed about $-2.7$, where \GELU stops being flat (see \autoref{fig:example}, bottom). In our simulation, across 400,000 sampled output locations the half-column sum never exceeded $-3.7$ and the largest per-location difference between the two programs was $5\times10^{-4}$.

To find such bugs efficiently we must shift our frame of reference: rather than checking all output locations for a single input, what if we could check a single output location for all inputs? Our bug is not specific to one output location: any location can be wrong given sufficiently positive input values. So we select an arbitrary location, say $(0,0,0,0)$, and construct symbolic expressions for the value of this output location in the reference and optimized implementations:
\[
  \mathsf{gelu}\bigl(S(64)\bigr) + b[0]
  \;\stackrel{?}{=}\;
  \mathsf{gelu}\bigl(S(32)\bigr) + b[0]
\]
where $S(H) = \sum_{h<H} \min_{c<16} M[0,c,h,0]$ and $M$ is the post-convolution tensor.
Since the shapes are statically known, a chosen location fixes the bounds of the summation and minimum, and an off-the-shelf SMT solver can find a witness to inequivalence.
We can even push additional constraints to the solver to find a counterexample that makes the difference visible above a given tolerance.

The natural question, then, is does this property generalize to other tensor programs, and how can we efficiently generate symbolic expressions for arbitrary tensor programs?
Next, we describe \dirigo, a system that lets us answer these questions.

\section{\dirigo}
\label{sec:dirigo}

\dirigo is a system for equivalence-checking of tensor programs based on \hypothesis that supports analysis of both high-level PyTorch operations and low-level CUDA kernels.
It is organized into four stages, as depicted in \autoref{fig:overview}: (1) \TRACE in which the trace of each program is extracted via runtime instrumentation, (2) \MERGE in which the traces are combined and optimized by discharging trivial equivalence obligations, (3) \EXECUTE in which the kernels and Torch operations are lowered to symbolic expressions, and (4) \CHECK in which the symbolic expressions are concretized for specific output tensor locations and passed to the Z3 SMT solver to look for counterexamples.

\subsection{\TRACE}
\label{sec:tracer}

\input{fig-trace-merge}

The first step is to extract a representation of the program.
We consider only programs that take tensors as inputs and produce tensors as output (possibly aliased from the inputs).
We also require that given the same input tensor shapes, the program exhibits the same control flow and intermediate tensor shapes regardless of the input values.
Further, our prototype assumes the only producers of tensors are Torch APIs (across Python and C++), and the only consumers are Torch APIs and CUDA runtime APIs, although this could extend to kernel DSLs like Triton and libraries such as JAX.

These assumptions are not particular to \dirigo: the compilation paths of the major tensor frameworks already impose them, so programs written to be compiled satisfy them as a matter of course.
JAX~\cite{jax2018} traces a function once on shape-only values and reuses the resulting trace for every input of that shape.
Using a Python conditional on a tensor value is a trace-time error (data-dependent control needs an explicit \texttt{lax.cond}).
PyTorch's \texttt{torch.compile}~\cite{pytorch2compile2024} takes the same position: TorchDynamo captures a graph specialized to the input shapes, graph-breaks or fails when Python control flow depends on tensor values, and requires explicit \texttt{torch.cond} where such dependence is intended.
Requiring the same of the programs we analyze therefore costs nothing for the large body of code that already targets these systems, including the KernelBench benchmark~\cite{kernelbench2025}.

With these assumptions in mind, the program's behavior is exactly the set of Torch operations and CUDA runtime events that are executed.
Intuitively, this is because:
\begin{enumerate}
  \item We disregard side-effects (beyond internal mutation), so the only data we care about is the output tensors.
  \item The only way for output tensors to be modified is via Torch operations and CUDA runtime events.
  \item The same set of operations must be executed regardless of the input values, since the control flow is data-independent.
\end{enumerate}

To extract these events, we use a combination of tricks:
for Python, \texttt{TorchDispatchMode} records all PyTorch invocations;
for C++, \texttt{at::RecordFunction} records all ATen operations (the C++ bindings to Torch);
for CUDA, we use an \texttt{LD\_PRELOAD} shim to interpose on the CUDA runtime APIs and record all kernel launches (along with the PTX blob), memory copies, and symbols.
Since Torch is the only producer of tensors, we can keep a live list of tensor pointer addresses to map kernel arguments to tensors.
Since Torch operations call the CUDA runtime internally, we suppress the shim during Torch calls.
The process is depicted in the top right of \autoref{fig:trace-merge}.

To verify that programs exhibit ``traceable behavior'', we run them multiple times with different tensors and check that the trace is the same. Running \texttt{torch.compile} could also be used to verify the desired behavior.

\subsection{\MERGE}
\label{sec:merger}

Once both purportedly equivalent programs are traced, we combine them into a single equivalence-checking obligation.
The first step is to convert them to SSA form to make reasoning about aliasing and mutation easier.
Then we match the input and output tensors of the two traces to encode the goal that the same inputs should produce the same outputs. This is illustrated in the bottom of \autoref{fig:trace-merge}.

Two further optimizations make equivalence-checking easier: if the two traces share a common prefix or suffix of Torch operations, we can trim it to reduce the operations replayed by the next stage, \EXECUTE.
This only preserves equivalence under certain conditions: the trimmed prefix must be surjective, and the suffix injective.
However, even when they don't meet these conditions, trimming does not change \dirigo's output, since the spurious counterexample found by \CHECK still produces equivalent results on the original programs, so \dirigo will not alert the user.
In practice we enable prefix trimming, since nearly all prefix operations are surjective, but not suffix trimming, since few suffix operations are injective.

\subsection{\EXECUTE}
\label{sec:executor}

The \EXECUTE stage is a core contribution of \dirigo, as it demonstrates a novel technique for applying symbolic execution to CUDA kernels, where methods designed for CPU code fail.
Intuitively, traditional CPU-focused symbolic executors consider each thread individually, but abstract over many executions of that thread, such as differing loop bounds and dynamic branching.
From this perspective, the GPU code we consider is much simpler as the loops and branches are static with respect to our symbolic variables.
Instead, our complexity comes from abstracting over the many threads executing the same code in parallel, and their interactions.
Concretely, we treat the thread ID and block ID (\tid and \bid), the only data threads can use to distinguish themselves, as symbolic values.
A single logical thread of symbolic execution then captures an unbounded number of physical GPU threads, and we recover a single thread's behavior by substituting concrete \tid and \bid.
The natural question, then, is how to handle interactions between threads, such as shared memory and warp-level reductions, which we address next.

\subsubsection{Setup}

The CUDA kernel symbolic execution problem statement is as follows: given a program text $P$ that reads a set of input tensors $\bar{x}$ and writes a set of output tensors $\bar{y}$, as well as a block count $b$ and a threads-per-block count $t$, we want to produce, for each output tensor $y \in \bar{y}$, a map $R_y : [0, |y|) \to \mathcal{S}$ from a flat index into $y$ to a symbolic expression over the inputs such that
\begin{align*}
  \forall \bar{x}, \bar{y}.\; [\![P]\!](\bar{x}, b, t) = \bar{y} \iff{} & \forall y \in \bar{y}.\; \forall i \in [0, |y|),\\
  & [\![R_y(i)]\!]_{\bar{x}} = y[i]
\end{align*}
where $[\![P]\!]$ is the semantics of the program $P$, $\mathcal{S}$ is the set of symbolic expressions over the elements of $\bar{x}$, and $[\![e]\!]_{\bar{x}}$ evaluates a symbolic expression $e$ by substituting the concrete elements of $\bar{x}$ for the corresponding symbolic variables.

We model integers as bitvectors, since bitwise operations are prevalent in GPU kernels, but floats as reals, since we deem precision errors as out of scope for this work.
This is both desirable, as optimized kernels may not be bitwise equivalent, and undesirable, as it doesn't let us reason about properties like numerical stability.

We accomplish this by tracking an environment that maps registers to symbolic expressions.
This allows us to easily execute instructions like \texttt{add r1, r2, r3} by looking up the symbolic expressions for \texttt{r2} and \texttt{r3} and producing a new symbolic expression for \texttt{r1}.
More complicated are expressions that involve the special registers \tid and \bid, which we now cover.

\input{fig-executor}

\subsubsection{Branching on \tid and \bid}

When execution of a kernel begins, one logical symbolic execution thread captures the behavior of all kernel threads, with constant scaling in the number of threads.
However, many kernels branch on \tid or \bid to for instance mask out certain threads, generally with a \texttt{setp} (``set predicate'') instruction and a predicated jump, such as
\begin{lstlisting}[style=inline,language=PTX]
setp.lt.u32 p1, %tid, 128
@!p1 bra label_1
\end{lstlisting}
Inserting a conditional into the symbolic expression to execute the \texttt{setp} line would leave us stuck at the branch, since we require a concrete value to determine the branch target (existing methods handle symbolic branch conditions, but concrete targets are much simpler).
Instead, at \texttt{setp} instructions that only depend on \tid and \bid, we split our logical symbolic execution thread into two, one with the \tid and \bid values satisfying the predicate and one with their complement (\autoref{fig:bundles-a}).
This way, when we arrive at the branch the predicate is a uniform concrete value.
We do not attempt to later coalesce these threads if their executions converge.
This could hypothetically explode the number of logical threads, but in practice branching is manageable: we observe a median of 3 logical threads per kernel and a 90th percentile of 15 across the kernels in our evaluation.
For \texttt{setp} instructions that depend on tensor values we do create a symbolic conditional, since our data-independence assumption guarantees the predicate is never used for branching.

\subsubsection{Shared memory and thread barriers}

Most high-performance kernels use shared memory to share intermediate results between threads in a block and eliminate redundant global memory reads.
For shared memory stores to be visible across threads, all threads must synchronize at a barrier, \texttt{\_\_syncthreads()} in CUDA and \texttt{bar.sync} in PTX. The code we must support looks like
\begin{lstlisting}[style=inline,language=C++]
smem[tid] = gmem_in[tid];
__syncthreads();
for (int s = 0; s < blockDim.x; s++)
  gmem_out[tid] += smem[s];
\end{lstlisting}
To capture this behavior, the execution context contains both a symbolic expression representing the current content of shared memory, and the set of pending writes to shared memory (suppose without loss of generality that there is exactly one shared memory region per kernel).
When a symbolic execution thread bundle reaches a barrier, it cannot step until all bundles have reached the barrier.
At that point, the existing shared memory content is updated with the pending writes, and the pending writes are cleared.
One situation complicates this:
a buggy program may race, reading shared memory over a pending write. However, not every read with pending writes is a race, since a thread may be reading a value it wrote itself (which can never race, by program order).
To solve this, we use an SMT solver to check for overlap between pending read and write sets at each barrier. \dirigo also performs out-of-bounds access checks for shared and global memory.

\subsubsection{Intra-Warp Reduction}
The shared memory example shows one way to perform a block-wide reduction, but reducing across 32 or fewer threads is faster with \emph{intra-warp primitives}, intrinsics that let threads in a warp communicate without shared memory.
A common example is \texttt{shfl.down}, which lets each thread read a value from another thread a fixed offset away.
The following snippet from the running example sums a value among 32 threads:
\begin{lstlisting}[style=inline,language=C++]
float val = sdata[tid];
for (int offset = 16; offset > 0; offset /= 2) {
  val += __shfl_down_sync(0xffffffff, val, offset);
}
\end{lstlisting}
On each iteration, each thread reads a value from another thread in the warp and adds it to its own.
By the CUDA specification, a thread reading from a lane outside the warp instead reads its own value.
Therefore, an implementation of \texttt{shfl.down r2, r1, $c$} in the symbolic executor with $\texttt{r1}=\alpha_{\tid}$ is
\[\texttt{r2}=\begin{cases}\alpha_{\tid+c} & \text{if } (\tid \bmod 32) + c < 32 \\ \alpha_{\tid} & \text{otherwise}\end{cases}\]
However, this rests on the faulty assumption that every lane of the warp is in the same thread bundle. After a split on \tid, lanes $0$--$15$ and lanes $16$--$31$ may be held by different logical threads with different expressions for \texttt{r1}, making the above expression incorrect.
Instead, we treat intra-warp primitives as a barrier, waiting for all threads to arrive before resolving \texttt{r2} for each lane.

We compute \texttt{r2} as follows: let a reader have partner lane $\pi(\tid)$, and for each member $S$ of the group we form the predicate $\mathit{valid}_S(\pi(\tid))$ over the \emph{reader's} \tid obtained by substituting $\pi(\tid)$ into $S$'s domain, and keep $S$ when that guard is satisfiable together with the reader's domain.
The valid members become the arms of a conditional expression, each substituting $\pi(\tid)$ into that member's value of the source register.
Writing $\alpha^{S}$ for the value of \texttt{r1} held by member $S$, the example above with $c=16$ becomes
\[\texttt{r2}=\begin{cases}
  \alpha^{S_0}_{\tid} & \text{if } (\tid \bmod 32) + 16 \ge 32 \\
  \alpha^{S_1}_{\tid+16} & \text{if } \mathit{valid}_{S_1}(\tid+16) \\
  \alpha^{S_2}_{\tid+16} & \text{if } \mathit{valid}_{S_2}(\tid+16) \\
  \vdots & \vdots
\end{cases}\]
where $S_0$ is the reader itself.
This is the fully general form, but in practice \dirigo detects when a warp is entirely contained in one bundle and uses the simpler formula above.

Finally, an intra-warp primitive may read from an exited thread, or from a thread beyond the end of the block when the block size is not a multiple of 32.
This is not itself an error and produces an undefined value, modeled as $\bot$, but if it reaches an observable sink---a memory write, a memory address, or a branch predicate---we flag a bug.
\autoref{fig:bundles-b} annotates some of the running example's PTX with the symbolic values produced by \EXECUTE, ending in the predicated write that only the bundle holding $\tid = 0$ performs.

\subsubsection{Torch Operations}
Unlike kernels, which we must consider at the PTX level, Torch operations have a pre-defined mapping to symbolic expressions.

\subsection{\CHECK}
\label{sec:checker}

\input{fig-checker}

After \EXECUTE, every operation in both programs has a symbolic description, which we must turn into equivalence conditions a solver can find counterexamples to.
Ideally we would generate one formula per program capturing its full behavior and ask the solver whether the two agree over entire output tensors.
But such formulas are dominated by conditionals on thread indices and bit-vector address arithmetic, on which SMT solvers like Z3~\cite{z3_2008} either time out or return unknown.
Instead, \CHECK takes advantage of \hypothesis to reduce the problem to equivalence at a single output location.

\mypara{Flipping the quantifiers.}
Let $A$ and $B$ be the two programs (the optimized and reference programs, respectively), $\bar{x}$ their (shared) input tensors, and $y$ an output tensor with index domain $\mathrm{dom}(y)$, and write $A(\bar{x})[\ell]$ for the value $A$ stores at location $\ell \in \mathrm{dom}(y)$.
Equivalence is the statement
\begin{equation*}
  \forall \ell \in \mathrm{dom}(y).\; \forall \bar{x}.\; A(\bar{x})[\ell] = B(\bar{x})[\ell].
\end{equation*}
Differential testing fixes an input $\bar{x}$ and checks every $\ell$ concretely.
\CHECK does the opposite: it fixes a concrete $\ell$ and asks a solver to decide the inner formula $\forall \bar{x}.\, A(\bar{x})[\ell] = B(\bar{x})[\ell]$ symbolically.
This is what makes the query tractable.
With $\ell$ concrete and shapes static, every index computation, thread-index conditional, and bit-vector operation evaluates to a constant, since memory access and control flow are data-independent by assumption.
What remains is a quantifier-free formula over real arithmetic and uninterpreted functions (one per transcendental, e.g.\ \texttt{tanh}), with conditionals only for data-dependent operations like \texttt{max} and clamping, which is much easier for Z3 to solve.

\mypara{Per-element equations.}
After \MERGE, each program is a sequence of operations $\mathit{op}_1, \ldots, \mathit{op}_m$ in SSA form.
For simplicity, assume $\mathit{op}_k$ writes exactly one tensor $T_k$ and reads only input tensors and tensors $T_j$ with $j < k$.
We treat every tensor element $T[\ell]$ as a distinct symbolic variable.
Abstractly, \EXECUTE produces a \emph{per-element equation} for each operation: a function $f_k$ mapping a concrete location $\ell \in \mathrm{dom}(T_k)$ to a symbolic expression over elements of earlier tensors, such that $T_k[\ell] = f_k(\ell)$.
For Torch operations, $f_k$ is given directly by the lowering of \autoref{sec:executor}; in the running example, $f_{\texttt{add}}(n,c,w) = G[n,w] + b[c]$.
For kernels, \EXECUTE instead produces a set of symbolic writes $(\pcond, a, v)$: under path condition $\pcond(\tid,\bid)$, a thread writes the value $v(\tid,\bid)$ to the address $a(\tid,\bid)$.
To obtain $f_k(\ell)$ we invert the write: we find the $(\tid,\bid)$ satisfying $a(\tid,\bid) = \ell \wedge \pcond(\tid,\bid)$ and substitute it into $v$.
Since $\ell$ is concrete and $a$ is affine in \tid and \bid for the kernels we encounter, this is a small integer problem.

\mypara{Back-projection.}
A concrete output location is resolved to an expression over the inputs alone by substituting per-element equations backwards through the program:
\begin{align*}
  \mathrm{res}(x[\ell]) &= x[\ell] \\
  \mathrm{res}(T_k[\ell]) &= f_k(\ell)\big[\,T_j[\ell'] \mapsto \mathrm{res}(T_j[\ell'])\,\big]
\end{align*}
where $x$ ranges over input tensors and the substitution is applied to every element $T_j[\ell']$ occurring in $f_k(\ell)$.
Every $\ell'$ that appears is itself concrete, so the recursion touches finitely many elements and terminates by the SSA ordering.
Applying $\mathrm{res}$ to $y[\ell]$ in each program yields $e_A$ and $e_B$, two expressions over the same input variables (\autoref{fig:checker}).

\mypara{Solving.}
The solver checks $e_A \! \neq \! e_B$.
An \unsat result means the location is equivalent: $\forall \bar{x}.\, A(\bar{x})[\ell] = B(\bar{x})[\ell]$ holds up to our modeling assumptions.
A \sat result yields a witness $\bar{x}^*$, a concrete input where the programs disagree at $\ell$.

\mypara{Enumeration and budgets.}
Repeating this process for every $\ell \in \mathrm{dom}(y)$ discharges the outer quantifier and verifies the programs in full, up to our modeling assumptions.
However, this is not always pragmatic.
\Hypothesis says we should expect to find a bug, if one exists, after only a few locations.
So where fast iteration matters, it is more useful to set a time budget and check as many locations as possible within it.
This lets a human kernel author iterate quickly (running \dirigo on each compilation) and an agentic system spend its limited compute budget effectively.
A budget therefore adds a third outcome, unknown, when it is exhausted before the solver has checked at least $n$ locations (in practice, we set $n=5$).
Finally, budgeted checking improves further with a good order for checking locations.
This heuristic can be quite rich, incorporating path conditions gathered during symbolic execution to catch pathological bugs that occur at only a few locations; we defer discussing it to \autoref{sec:impl-heuristic}.
\section{Implementation}
\label{sec:impl}

We now discuss \dirigo implementation details that matter for effectiveness in practice.

\subsection{PTX-Level Transcendental Rollup}
Like most GPU compilers, the CUDA compiler \texttt{nvcc} does not compute exact results for transcendental functions like \texttt{tanh} and \texttt{log}, which activation functions commonly use.
Instead it emits an approximation: a polynomial, or a piecewise method that chooses between two approximations based on the magnitude of the input.
This is problematic for \dirigo: it creates data-dependent control flow where none existed and obscures the source program's mathematical meaning.
To solve this, \dirigo pre-processes the PTX code to recognize \texttt{nvcc} approximation idioms and convert them to a single pseudo-instruction that represents the mathematical function called in the source.
In practice, this is not too challenging: in the entire \archive, the only transcendental functions are \texttt{exp}, \texttt{log}, \texttt{log1p}, \texttt{tanh}, \texttt{erf}, and \texttt{pow}. Including fast math variants, \dirigo needs only eleven pattern matchers to handle these.

\subsection{Transcendental Approximation Axioms}
\label{sec:impl-axioms}
Commonly, a programmer might optimize a tensor program by replacing a slower, more mathematically precise transcendental function implementation with a faster, less accurate approximation.
\dirigo includes a pluggable set of axioms that can be applied to the equivalence-checking phase to allow the user to specify which approximations are acceptable.
We distinguish two kinds of rules.

\emph{Sound identities} are two implementations of a single mathematical function and are always on.
They are needed because Torch operations and kernels produce different symbolic expressions for activation functions: for example, kernels would produce $1/(1+e^{-x})$ to represent \texttt{sigmoid}, and Torch operations would produce an uninterpreted function $\mathsf{sigmoid}$.
\dirigo runs a pass over both final expressions to normalize uninterpreted functions to a common vocabulary.

\emph{Approximation equivalences} are approximations that can optionally be marked as equivalent to enable a more permissive notion of program equivalence.
The \archive needs only one: $\mathsf{gelu}_{\mathsf{erf}} = \mathsf{gelu}_{\mathsf{tanh}}$, which equates the \texttt{tanh} approximation of \GELU with the exact \texttt{erf}-based \GELU.
Similarly, we can optionally assume that fast-math variants are equivalent to their non-fast-math counterparts.
Unless otherwise specified, evaluations of \dirigo assume both these equivalences are enabled.

\subsection{Search Heuristic}
\label{sec:impl-heuristic}
Instead of exploring output locations randomly, the white-box nature of \dirigo's symbolic execution, combined with priors about where bugs manifest in deep learning kernels, allows for richer heuristics.
From the path conditions that \EXECUTE produced (each of which has its own symbolic write address and value), we rank a prefix of candidate locations in three tiers:
\begin{enumerate}
  \item \textbf{Corners.} For every kernel launch, the locations written by the first and last thread of the first and last block, $(\tid,\bid) \in \{0, t{-}1\} \times \{0, b{-}1\}$, as well as the warp-boundary lanes $\tid \in \{31, 32\}$ of block 0, where warp-reductions might contain bugs. Later kernels are ranked above earlier kernels.
  \item \textbf{Path-condition coverage.} For each unique thread bundle write not already covered by corner checking, we use Z3 to find a satisfying assignment of \tid and \bid that falls within that write's path condition. This ensures coverage of piecewise behavior that only exists in small portions of the output tensor.
  \item \textbf{Tensor boundaries.} The first and last element of every output tensor.
\end{enumerate}
After these, we fall back to a random permutation of the unchecked locations.
We evaluate this heuristic against a baseline sequential order in \autoref{sec:eval-rq2}.

\subsection{Parallelization}
\dirigo's structure exposes natural parallelism: \EXECUTE analyzes trace operations independently, so symbolic execution can be parallelized across operations, and \CHECK analyzes output locations independently, so checking can be parallelized across locations.
We evaluate the effect of location-level parallelism on \dirigo's ability to fully check all locations for equivalence in \autoref{sec:eval-rq4}.

\subsection{Modeling Gaps and Limitations}
\label{sec:impl-limitations}

\dirigo models floating point values as reals, a common abstraction used with SMT solvers \cite{volta2025}. This masks floating-point behavior like non-associativity, and means that \dirigo cannot find bugs that manifest through insufficient precision. Supporting floating-point arithmetic is beyond the scope of this work but an interesting open problem: work on hardening \kernelbench shows that global absolute and relative tolerances cannot separate acceptable approximation from numerical error~\cite{sakana_robust2025,metr2025}.

\dirigo only models a subset of PTX instructions. That subset is enough to run the majority of the \archive programs, but with more engineering effort we could incorporate the remaining missing operators like \texttt{\_\_syncwarp} and \texttt{\_\_ballot\_sync} into \dirigo's design.

\subsection{Counterexample Verification}
\label{sec:impl-counterexample}

When \dirigo finds an inequivalence, it appends constraints to the SMT query to find a counterexample that produces large differences between the two programs, and uses PyTorch's auto-differentiation to invert witnesses back through any prefix operations \MERGE trimmed. This allows us to confirm the counterexample on the real programs, avoiding any modeling limitations.

\section{Evaluation}
\label{sec:eval}

We evaluate both the validity of \hypothesis and its effectiveness (via \dirigo) as a practical bug-finding tool by answering the following questions:

\noindent
\textbf{RQ1:} Does searching across output locations instead of input values discover more bugs?

\noindent
\textbf{RQ2:} Among buggy kernels, how many locations do bugs appear at?

\noindent
\textbf{RQ3:} Among buggy kernels, how quickly does output location search discover the bugs?

\noindent
\textbf{RQ4:} Can \dirigo fully verify kernels by checking all locations?

\subsection{Setup}

\input{fig-corpus}

We evaluate \dirigo on the \archive, one of the only large, public collections of optimized KernelBench implementations.
We focus on KernelBench level-2, since it has the complexity of multiple operators (unlike level-1) while still having many tasks (unlike level-3).
KernelBench level-2 programs consist of 3-6 Torch operators and the \archive versions are C++ functions calling Torch functions and custom CUDA kernels.
Since we seek bugs that differential testing misses, we consider only the 6,988 \archive programs it marks correct.
All experiments ran on a workstation with two Intel Xeon Gold 6248 CPUs (20 physical cores each), 754 GB of DRAM, and NVIDIA GeForce RTX 2080 Ti GPUs (11 GB) with CUDA 12.5 and driver 555.42.02.
Unless otherwise noted, \dirigo runs with a 2 minute timeout on \EXECUTE and a 4 minute timeout on \CHECK.
An overview of the dataset's key features is in \autoref{fig:corpus}.
Note that most programs have a single kernel, are approximately 100 lines, exhibit very little thread-splitting, and use shared memory and at least one torch operation.

\subsection{RQ1: Does \hypothesis discover more bugs?}
\label{sec:eval-rq1}
\input{tab-bug-categories}

To gauge the validity of \hypothesis, we run \dirigo across the 6,988 \archive level-2 kernels that differential testing marks correct.
Over these programs, \dirigo reports four outcomes:
\begin{itemize}
  \item \textbf{Buggy (600, 8.6\%).} Which shows that differential testing alone is not enough.
  \item \textbf{Checked Locations Correct (3,341).} Every output location \dirigo reached within its budget was proved equivalent to the reference, with at least 5 locations checked. While this is evidence of correctness, it is not a guarantee.
  \item \textbf{Timeout (1,175).} The 120\,s \EXECUTE or 240\,s \CHECK budget expired before a verdict.
  \item \textbf{Unsupported (1,872).} Programs that either violate \dirigo's assumptions or use features absent from our prototype, such as \texttt{\_\_syncwarp}.
\end{itemize}
A taxonomy of the types of bugs found is shown in \autoref{tab:bug-categories}.
374 are reported by \EXECUTE\ as violations of the CUDA programming model (shared-memory races, shuffles whose partner lane has exited, out-of-bounds and uninitialized reads), and 226 by \CHECK\ as inequivalences, which are proved with a concrete witness that is replayed on the original program and produces differences well above tolerances.
Requiring a verified counterexample for every reported bug makes \dirigo complete (it only reports real bugs) at the cost of soundness (it may miss bugs when no convincing counterexample is found).
When \CHECK\ finds a bug with no verifiable counterexample, \dirigo cannot tell whether the refinement heuristic simply failed, or the inequivalence is spurious and exposes a gap in \dirigo's modeling assumptions.
On the \archive, \dirigo produces 221 such unverifiable inequivalence results (over 32 tasks).

Note that the counts are dominated by certain tasks: the \archive\ contains many near-duplicate optimizations of each task, so the 600 buggy programs span 68 of the 100 level-2 tasks, and we report both.

For each bug category, \autoref{tab:bug-categories} also shows how many of those bugs were caught by Nvidia's \texttt{compute-sanitizer}. \dirigo catches many low-level safety bugs that the sanitizers miss, in addition to higher-level bugs outside their scope.

\subsection{RQ2: Are bugs dense in output-space?}
\label{sec:eval-rq2}
\input{fig-eval-prevalence}

Of the 600 bugs we find, we filter to the 226 found at the per-location checking stage rather than during symbolic execution, and ask how many output locations reveal each bug.
For each, we check locations in random order for 600\,s and extrapolate the fraction that differ to the total number of buggy locations. The results are shown in \autoref{fig:eval-prevalence}.
In all but 13 cases, the bugs are present at every location sampled. Consistent with this, 218 of the 226 bugs are exposed by the very first location \dirigo samples. This provides strong empirical evidence of the validity of \hypothesis\ on this dataset.
Further, the two bugs with the lowest prevalences (0.003 and 0.001) were found by the heuristic search after 32 and 1,289 locations respectively, whereas the sequential order checked 668 and 9,083 locations of the same programs without reaching either bug.

\subsection{RQ3: Is output-space search wall-clock efficient?}
\label{sec:eval-rq3}
\input{fig-eval-time}

While confidence that bugs are dense in output-space is useful, we also want to know whether \dirigo can sit inside a developer's loop as an interactive bug-finder.
Across the 600 faulty programs, we track the time from the start of \TRACE to the bug report.
The results are shown in \autoref{fig:eval-time}.
93.2\% of bugs are found within the first minute and 97.3\% within two, evidence that \dirigo could be used interactively.
\EXECUTE completes quickly, so those bugs are found early, whereas \CHECK continues indefinitely and finds a long tail.
The time to a bug is dominated by \TRACE, which adds a median of 18\,s per program (the reference and optimized programs are traced concurrently). Counting only the search itself, the median time to a bug is 1.9\,s and 95.7\% of bugs are found within a minute.
\EXECUTE is cheap for most programs, with a median of 0.6\,s and 10\,s at the 90th percentile.

\subsection{RQ4: Is full verification of programs feasible?}
\label{sec:eval-rq4}
\input{fig-eval-fullverif}

We measure \dirigo over a 120\,s window with 40 threads and extrapolate to find the latency of full verification of all output locations.
To check these numbers, we also run full verification for every program estimated to take under 60 minutes.
The results are presented in \autoref{fig:eval-fullverif}.
While 58\% are projected to be fully verifiable in under 60 minutes, full verification is generally hard: some kernels are estimated at up to 61,440 hours (7.0 years). The extrapolation is generally accurate, with a median ratio of predicted-to-actual of 1.03.

\subsection{Comparison to Volta}
\label{sec:volta-comparo}

Volta \cite{volta2025}, a system for automated kernel equivalence-checking, is the most similar work to \dirigo: both use symbolic execution to accumulate equivalence conditions as SMT formulas passed to Z3. However, its limitations inhibit applicability to tasks like KernelBench. Volta only supports what it terms ``Structured CTA'' kernels, limited to data-independent control flow and memory access, constraints \dirigo shares. Volta is also restricted to checking two CUDA kernels against each other, and cannot check a kernel against a PyTorch operator. Additionally, Volta requires the baseline and optimized kernels to have identical grids, and compares one CTA (thread block) in the baseline to the corresponding CTA in the optimized kernel, checking that the output locations written by each CTA are equivalent. Checking one CTA's outputs is consistent with \hypothesis, but requiring CTA alignment precludes changing grid dimensions and kernel fusion. In multi-kernel programs, analyzing even one thread block of the last kernel commonly requires many thread blocks of the prior kernel, as with reductions that collapse many inputs to one output.

Despite this, we explore what applying Volta to the \archive might look like. The smallest change would be support for checking kernels against individual PyTorch operations, similar to the kernel-to-kernel checking Volta already supports. With it, one could check the subset of the \archive where the agent optimizes each reference PyTorch operation into its own kernel: approximately 242 entries out of 6,988. Volta is not open source, so unfortunately we cannot run it on those entries.

\section{Related Work}
\label{sec:related}

A large body of work verifies CPU programs via symbolic execution and SMT solving, but for brevity we limit our discussion to techniques targeting GPU programs.

Work on \textbf{GPU equivalence-checking} is most similar to \dirigo: it establishes the correctness of an optimized GPU kernel by showing equivalence to a trusted baseline. Differential testing is the basic but popular approach, though it samples very sparsely from a large input space and misses many bugs. Volta \cite{volta2025}, which we compare to in \autoref{sec:volta-comparo}, adopts a more principled approach and uses SMT solving to prove equivalence, though it requires the baseline kernel be written in CUDA and does not support kernel fusion. ProofWright \cite{chatterjee2026proofwright} can prove equivalence between PyTorch and CUDA code via LLM-written Rocq proofs, but currently only produces proofs for a subset of \kernelbench level-1 kernels. Likewise Kuiper, a verified GPU programming framework~\cite{martinez2026kuiper}, has implemented \kernelbench level-1 kernels~\cite{martinez2026nextfrontier}. These approaches require the kernel to be written or proved in a proof assistant, whereas \dirigo checks existing, unannotated CUDA against the PyTorch reference directly.

\textbf{GPU symbolic execution} systems \cite{gklee2012,collingbourne2014kleecl,collingbourne2012kleecl2,wu2020simulee,farooqui2014lynx-symex} detect a variety of correctness and performance bugs. \dirigo uses symbolic execution not only for bug detection but also to enable equivalence-checking. \dirigo's per-location concretization step (\autoref{sec:checker}) in particular takes inspiration from concolic execution systems \cite{godefroid2005dart,gklee2012}.

Many \textbf{GPU static analyses} \cite{cogumbreiro2024fmsd,faial2021cav,li2010pug,pereira2016esbmcgpu,alur2017gpudrano,alur2022uncoalesced,alur2018blocksize} detect data races, barrier divergence and other GPU correctness and performance bugs. The state-of-the-art FaialAA system \cite{liew2024faialaa} verifies that a kernel is DRF and identifies any data races it contains, so long as the kernel is control-independent (CI) and data-independent (DI). Such CIDI kernels are similar to the kernels that \dirigo supports, as they both lack data-dependent control flow and indirect memory accesses.

\section{Conclusion}
We have presented \hypothesis and used \dirigo to validate it on a large corpus of optimized deep learning kernels. Moreover, while many kernel bugs are simple safety violations, there is a long tail of complicated correctness violations that differential testing struggles to catch. \dirigo provides rich feedback and finds bugs quickly, accelerating both human and agentic kernel authors.

\bibliographystyle{ACM-Reference-Format}
\bibliography{references}

\end{document}

%% file: preamble-common.tex
\usepackage{xcolor}
\usepackage{tikz}
\usetikzlibrary{arrows.meta,positioning,shapes.geometric,shapes.misc,fit,calc,backgrounds,matrix,decorations.pathreplacing,chains,scopes}
\usepackage{pgfplots}
\pgfplotsset{compat=1.18}
\usepgfplotslibrary{statistics}
\usepackage{booktabs}
\usepackage{array}
\usepackage{mathpartir}
\usepackage{listings}
\usepackage{subcaption}
\usepackage{pifont}
\usepackage{xspace}
\usepackage{microtype}

\hypersetup{citecolor=blue,colorlinks=true,linkcolor=blue}

\newcommand{\dirigo}{\textsc{Dirigo}\xspace}

\newcommand{\kernelbench}{KernelBench\xspace}
\newcommand{\archive}{AI CUDA Engineer dataset\xspace}

\newcommand{\stagename}[1]{\textsc{#1}\xspace}
\newcommand{\TRACE}{\stagename{Trace}}
\newcommand{\MERGE}{\stagename{Merge}}
\newcommand{\EXECUTE}{\stagename{Execute}}
\newcommand{\CHECK}{\stagename{Check}}

\newcommand{\tid}{\ensuremath{\mathit{tid}}\xspace}
\newcommand{\bid}{\ensuremath{\mathit{bid}}\xspace}

\newcommand{\wset}{\ensuremath{\mathcal{W}}}

\newcommand{\pcond}{\ensuremath{\phi}}

\newcommand{\botval}{\ensuremath{\bot}}
\newcommand{\ite}[3]{\ensuremath{\mathsf{ite}(#1,\,#2,\,#3)}}

\newcommand{\sat}{\textsc{sat}\xspace}
\newcommand{\unsat}{\textsc{unsat}\xspace}

\definecolor{cA}{HTML}{1B6CA8}
\definecolor{cB}{HTML}{D1495B}
\definecolor{cC}{HTML}{2E8B57}
\definecolor{cD}{HTML}{E9A03B}
\definecolor{cE}{HTML}{7B4EA3}
\definecolor{cF}{HTML}{4A4A4A}
\colorlet{cAl}{cA!15}\colorlet{cBl}{cB!15}\colorlet{cCl}{cC!15}
\colorlet{cDl}{cD!20}\colorlet{cEl}{cE!15}\colorlet{cFl}{cF!12}
\colorlet{zonec}{cA!13}
\definecolor{dummyred}{HTML}{C0392B}

\tikzset{
  every node/.style={font=\footnotesize},
  stage/.style={draw, rounded corners=2pt, thick, align=center, minimum height=2.2em, inner sep=4pt, fill=white},
  artifact/.style={draw, align=center, inner sep=3pt, fill=cFl, font=\footnotesize},
  tensorbox/.style={draw=cC, sharp corners, fill=cCl, align=center, inner sep=3pt, font=\footnotesize},
  iotensor/.style={tensorbox, very thick, fill=cC!28},
  torchop/.style={draw=cA, thick, rounded corners=5pt, fill=cAl, align=center, inner sep=3pt, font=\footnotesize\ttfamily},
  cudaop/.style={draw=cB, thick, rounded corners=5pt, fill=cBl, align=center, inner sep=3pt, font=\footnotesize\ttfamily},
  flow/.style={-{Latex[length=2mm]}, thick},
  dflow/.style={-{Latex[length=1.6mm]}, semithick, cF},
  dummy/.style={font=\footnotesize\bfseries, text=dummyred},
}

\lstdefinelanguage{PTX}{
  morekeywords={ld,st,mov,setp,bra,shfl,sync,down,bfly,min,max,add,mul,mad,and,or,cvta,ret,param,global,f32,s32,u32,u64,b32,pred,entry,reg,visible},
  sensitive=true, morecomment=[l]{//},
}
\lstdefinestyle{code}{
  basicstyle=\footnotesize\ttfamily, keywordstyle=\bfseries, commentstyle=\color{cF}\itshape,
  columns=fullflexible, keepspaces=true, breaklines=false, showstringspaces=false,
  numbers=none, xleftmargin=0pt, frame=none, escapeinside={(*@}{@*)},
}
\lstdefinestyle{inline}{
  style=code, xleftmargin=1.5em,
  aboveskip=0.5\baselineskip, belowskip=0.5\baselineskip,
}

\newcommand{\GELU}{\texttt{GELU}\xspace}
\newcommand{\hypothesis}{the Output-Space Hypothesis\xspace}
\newcommand{\Hypothesis}{The Output-Space Hypothesis\xspace}

\newcommand{\mypara}[1]{\par\noindent\textbf{\textit{#1}}}

%% file: fig-intro.tex
\begin{figure}[t]
\centering
\begin{tikzpicture}[x=1cm,y=1cm]
  \pgfmathsetseed{2026}
  \definecolor{mutedred}{HTML}{D88080}

  \pgfmathsetmacro{\tc}{0.09}
  \pgfmathsetmacro{\tg}{0.022}
  \pgfmathsetmacro{\tstep}{\tc+\tg}
  \pgfmathsetmacro{\tw}{3*\tstep-\tg}
  \newcommand{\thumb}[3]{%
    \begin{scope}[shift={(#1,#2)}]
      \foreach \i in {0,1,2} \foreach \j in {0,1,2} {
        \pgfmathrandominteger{\g}{20}{80}
        \ifnum#3=2 \fill[mutedred] (\i*\tstep,\j*\tstep) rectangle ++(\tc,\tc);
        \else \ifnum#3=1 \fill[cA!\g!white] (\i*\tstep,\j*\tstep) rectangle ++(\tc,\tc);
        \else \fill[gray!\g!white] (\i*\tstep,\j*\tstep) rectangle ++(\tc,\tc); \fi\fi
      }
      \ifnum#3=1 \draw[cA,thick] (-0.045,-0.045) rectangle (\tw+0.045,\tw+0.045); \fi
    \end{scope}}
  \newcommand{\thumbD}[3]{%
    \begin{scope}[shift={(#1,#2)}]
      \foreach \i in {0,1,2} \foreach \j in {0,1,2} {
        \pgfmathrandominteger{\g}{20}{80}
        \ifnum\j=1
          \ifnum#3=2 \fill[cB] (\i*\tstep,\j*\tstep) rectangle ++(\tc,\tc);
          \else \fill[cA!\g!white!80!cA] (\i*\tstep,\j*\tstep) rectangle ++(\tc,\tc); \fi
        \else
          \ifnum#3=2 \fill[mutedred!35!white] (\i*\tstep,\j*\tstep) rectangle ++(\tc,\tc);
          \else \fill[gray!\g!white] (\i*\tstep,\j*\tstep) rectangle ++(\tc,\tc); \fi
        \fi
      }
      \ifnum#3=2 \draw[cB,thick] (-0.045,\tstep-0.045) rectangle (\tw+0.045,\tstep+\tc+0.045);
      \else \draw[cA,thick] (-0.045,\tstep-0.045) rectangle (\tw+0.045,\tstep+\tc+0.045); \fi
    \end{scope}}

  \pgfmathsetmacro{\regw}{2.70}
  \pgfmathsetmacro{\regh}{1.85}
  \newcommand{\region}[1]{\draw[#1] (0,0) rectangle (\regw,\regh);}

  \def\positions{0.18/1.38/0,0.92/1.30/2,1.72/1.42/1,0.52/0.82/1,1.32/0.78/0,2.20/0.92/1,0.22/0.22/1,1.02/0.20/0,1.80/0.26/0,2.28/0.34/0}

  \begin{scope}
    \node[font=\bfseries,align=center,anchor=south] at (\regw/2,1.95) {Differential\\Testing};
    \region{gray!60,thick,fill=gray!6}
    \foreach \px/\py/\m in \positions { \thumb{\px}{\py}{\m} }
  \end{scope}

  \pgfmathsetmacro{\dx}{3.00}
  \begin{scope}[shift={(\dx,0)}]
    \node[font=\bfseries,align=center,anchor=south] at (\regw/2,1.95) {\dirigo\\(ours)};
    \region{gray!60,thick,fill=gray!6}
    \foreach \px/\py/\m in \positions { \thumbD{\px}{\py}{\m} }
    \draw[black,thick] (2.20-0.045,0.92-0.045) rectangle ++(\tw+0.09,\tw+0.09);
  \end{scope}

  \node[font=\small\sffamily,gray!80!black,align=center] at (\dx/2+\regw/2,-0.25)
    {space of all possible input tensors};

  \pgfmathsetmacro{\zs}{0.19}
  \pgfmathsetmacro{\zg}{0.045}
  \pgfmathsetmacro{\zstep}{\zs+\zg}
  \pgfmathsetmacro{\hx}{\dx+2.20+\tw+0.045}
  \pgfmathsetmacro{\zoomX}{\hx+0.40}
  \pgfmathsetmacro{\zoomY}{0.32}
  \pgfmathsetmacro{\outX}{\zoomX+3*\zstep+0.62}
  \draw[black,dashed,semithick] (\hx,0.92+\tw+0.045) -- (\zoomX,\zoomY+3*\zstep-\zg);
  \draw[black,dashed,semithick] (\hx,0.92-0.045) -- (\zoomX,\zoomY);
  \node[align=center,font=\footnotesize\sffamily] at (\zoomX+1.5*\zstep-0.5*\zg,\zoomY+3*\zstep+0.52) {input\\tensor};
  \foreach \i in {0,...,2} \foreach \j in {0,...,2} {
    \pgfmathsetmacro{\cx}{\zoomX+\i*\zstep}
    \pgfmathsetmacro{\cy}{\zoomY+\j*\zstep}
    \ifnum\j=1
      \pgfmathrandominteger{\rb}{35}{95}
      \fill[cA!\rb!white,rounded corners=2pt] (\cx,\cy) rectangle ++(\zs,\zs);
      \coordinate (in-top-\i) at (\cx+0.5*\zs,\cy+\zs);
    \else
      \pgfmathrandominteger{\rg}{20}{85}
      \fill[gray!\rg!white,rounded corners=2pt] (\cx,\cy) rectangle ++(\zs,\zs);
    \fi
  }
  \node[align=center,font=\footnotesize\sffamily] at (\outX+0.5*\zs,\zoomY+3*\zstep+0.52) {output\\tensor};
  \foreach \j in {0,...,2} {
    \pgfmathsetmacro{\cy}{\zoomY+\j*\zstep}
    \ifnum\j=1
      \fill[cE!80!white,rounded corners=2pt] (\outX,\cy) rectangle ++(\zs,\zs);
      \coordinate (out-top) at (\outX+0.5*\zs,\cy+\zs);
    \else
      \pgfmathrandominteger{\rg}{20}{85}
      \fill[gray!\rg!white,rounded corners=2pt] (\outX,\cy) rectangle ++(\zs,\zs);
    \fi
  }
  \foreach \i in {0,...,2} { \draw[->,>=stealth,cA,thick] (in-top-\i) to[out=40,in=140] (out-top); }
\end{tikzpicture}
\caption{Differential testing (left) samples sparsely (blue squares) from a large input space, often missing bugs (in red). \dirigo starts from one output location $\ell$ (purple) and works backwards to find the input locations (blue) that affect $\ell$, checking them across all possible inputs.}
\label{fig:intro}
\end{figure}
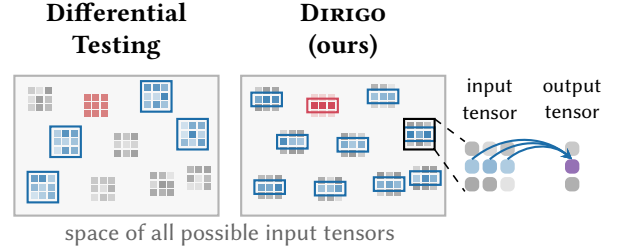

%% file: fig-example.tex
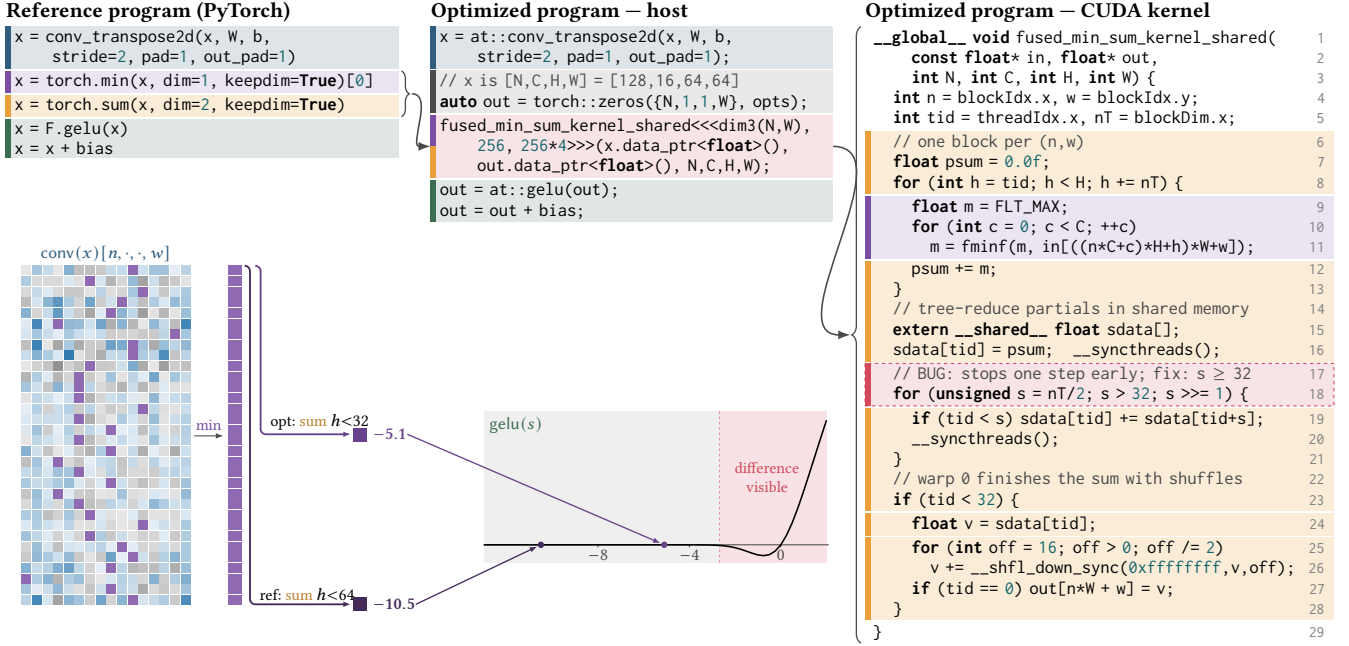
\begin{figure*}[th!]
\centering
\providecommand{\xin}[1]{\hspace*{#1\dimexpr 3.5pt\relax}}
\providecolor{synlit}{HTML}{12666B}
\colorlet{mconv}{cA!55!cF}\colorlet{mconvl}{cA!40!cF!16!white}
\colorlet{msuf}{cC!55!cF}\colorlet{msufl}{cC!40!cF!16!white}
\providecommand{\sykw}[1]{\textbf{#1}}
\providecommand{\sycm}[1]{\textcolor{cF}{#1}}%
\providecommand{\sylit}[1]{\textcolor{synlit}{#1}}
\providecommand{\lno}[1]{\hspace{0pt plus 1fill}{\tiny\color{cF!70}#1}}
\providecommand{\cv}[1]{\textcolor{mconv}{#1}}
\providecommand{\mn}[1]{\textcolor{cE}{#1}}
\providecommand{\sm}[1]{\textcolor{cD!80!black}{#1}}
\providecommand{\gl}[1]{\textcolor{msuf}{#1}}
\begin{tikzpicture}[
  x=1pt, y=1pt,
  ln/.style={anchor=north west, outer sep=0pt, inner xsep=3.5pt, inner ysep=1.2pt,
             font=\linespread{0.95}\scriptsize\ttfamily\selectfont, align=left},
  ca/.style={ln, text width=141pt},
  cb/.style={ln, text width=144.5pt},
  cc/.style={ln, text width=169.5pt},
  bar/.style={path picture={\fill[#1] (path picture bounding box.north west)
              rectangle ([xshift=2pt]path picture bounding box.south west);}},
  bar2/.style 2 args={path picture={%
    \fill[#1] (path picture bounding box.north west)
              rectangle ([xshift=2pt]path picture bounding box.west);
    \fill[#2] (path picture bounding box.west)
              rectangle ([xshift=2pt]path picture bounding box.south west);}},
  bug/.style={fill=cBl, bar=cB, draw=cB, line width=0.4pt,
              dash pattern=on 1.3pt off 1.3pt},
  hdr/.style={anchor=south west, inner sep=0pt, font=\footnotesize\bfseries},
  lbl/.style={font=\scriptsize\itshape, text=cF, inner sep=1pt},
  node distance=0.8pt,
]
\node[ca, fill=mconvl, bar=mconv] (A1) at (0,0)
  {x = conv\_transpose2d(x, W, b,\\
   \xin{4}stride=\sylit{2}, pad=\sylit{1}, out\_pad=\sylit{1})};
\node[ca, fill=cEl, bar=cE, below=of A1] (A2)
  {x = torch.min(x, dim=\sylit{1}, keepdim=\sykw{True})[\sylit{0}]};
\node[ca, fill=cDl, bar=cD, below=of A2] (A3)
  {x = torch.sum(x, dim=\sylit{2}, keepdim=\sykw{True})};
\node[ca, fill=msufl, bar=msuf, below=of A3] (A4)
  {x = F.gelu(x)\\
   x = x + bias};
\node[hdr] at ([yshift=1.5pt]A1.north west) {Reference program (PyTorch)};

\node[cb, fill=mconvl, bar=mconv] (B1) at (160,0)
  {x = at::conv\_transpose2d(x, W, b,\\
   \xin{4}stride=\sylit{2}, pad=\sylit{1}, out\_pad=\sylit{1});};
\node[cb, fill=cFl, bar=cF, below=of B1] (B2)
  {\sycm{// x is [N,C,H,W] = [128,16,64,64]}\\
   \sykw{auto} out = torch::zeros(\{N,\sylit{1},\sylit{1},W\}, opts);};
\node[cb, fill=cBl, bar2={cE}{cD}, below=of B2] (B3)
  {fused\_min\_sum\_kernel\_shared<{}<{}<dim3(N,W),\\
   \xin{4}\sylit{256}, \sylit{256}*\sylit{4}>{}>{}>(x.data\_ptr<\sykw{float}>(),\\
   \xin{4}out.data\_ptr<\sykw{float}>(), N,C,H,W);};
\node[cb, fill=msufl, bar=msuf, below=of B3] (B4)
  {out = at::gelu(out);\\
   out = out + bias;};
\node[hdr] at ([yshift=1.5pt]B1.north west) {Optimized program --- host};

\node[cc] (K0) at (323.5,0)
  {\sykw{\_\_global\_\_} \sykw{void} fused\_min\_sum\_kernel\_shared(\lno{1}\\
   \xin{4}\sykw{const} \sykw{float}* in, \sykw{float}* out,\lno{2}\\
   \xin{4}\sykw{int} N, \sykw{int} C, \sykw{int} H, \sykw{int} W) \{\lno{3}\\
   \xin{2}\sykw{int} n = blockIdx.x, w = blockIdx.y;\lno{4}\\
   \xin{2}\sykw{int} tid = threadIdx.x, nT = blockDim.x;\lno{5}};
\node[cc, fill=cDl, bar=cD, below=of K0] (K1)
  {\xin{2}\sycm{// one block per (n,w)}\lno{6}\\
   \xin{2}\sykw{float} psum = \sylit{0.0f};\lno{7}\\
   \xin{2}\sykw{for} (\sykw{int} h = tid; h < H; h += nT) \{\lno{8}};
\node[cc, fill=cEl, bar=cE, below=of K1] (K2)
  {\xin{4}\sykw{float} m = FLT\_MAX;\lno{9}\\
   \xin{4}\sykw{for} (\sykw{int} c = \sylit{0}; c < C; ++c)\lno{10}\\
   \xin{6}m = fminf(m, in[((n*C+c)*H+h)*W+w]);\lno{11}};
\node[cc, fill=cDl, bar=cD, below=of K2] (K3)
  {\xin{4}psum += m;\lno{12}\\
   \xin{2}\}\lno{13}\\
   \xin{2}\sycm{// tree-reduce partials in shared memory}\lno{14}\\
   \xin{2}\sykw{extern} \sykw{\_\_shared\_\_} \sykw{float} sdata[];\lno{15}\\
   \xin{2}sdata[tid] = psum;\xin{2}\_\_syncthreads();\lno{16}};
\node[cc, bug, below=of K3] (K4)
  {\xin{2}\sycm{// BUG: stops one step early; fix: s $\geq$ 32}\lno{17}\\
   \xin{2}\sykw{for} (\sykw{unsigned} s = nT/\sylit{2}; s > \sylit{32}; s >{}>= \sylit{1}) \{\lno{18}};
\node[cc, fill=cDl, bar=cD, below=of K4] (K5)
  {\xin{4}\sykw{if} (tid < s) sdata[tid] += sdata[tid+s];\lno{19}\\
   \xin{4}\_\_syncthreads();\lno{20}\\
   \xin{2}\}\lno{21}\\
   \xin{2}\sycm{// warp 0 finishes the sum with shuffles}\lno{22}\\
   \xin{2}\sykw{if} (tid < \sylit{32}) \{\lno{23}};
\node[cc, fill=cDl, bar=cD, below=of K5] (K6)
  {\xin{4}\sykw{float} v = sdata[tid];\lno{24}};
\node[cc, fill=cDl, bar=cD, below=of K6] (K7)
  {\xin{4}\sykw{for} (\sykw{int} off = \sylit{16}; off > \sylit{0}; off /= \sylit{2})\lno{25}\\
   \xin{6}v += \_\_shfl\_down\_sync(\sylit{0xffffffff},v,off);\lno{26}\\
   \xin{4}\sykw{if} (tid == \sylit{0}) out[n*W + w] = v;\lno{27}\\
   \xin{2}\}\lno{28}};
\node[cc, below=of K7] (K8) {\}\lno{29}};
\node[hdr] at ([yshift=1.5pt]K0.north west) {Optimized program --- CUDA kernel};

\draw[cF, line width=0.5pt, decorate,
      decoration={brace, amplitude=3pt, raise=1pt}]
  (A2.north east) -- (A3.south east);
\coordinate (bt) at ($(A2.north east)!0.5!(A3.south east)+(5pt,0)$);
\draw[dflow] (bt) to[out=0,in=180] (B3.west);
\draw[cF, line width=0.5pt, decorate,
      decoration={brace, mirror, amplitude=3.5pt, raise=1.5pt}]
  (K0.north west) -- (K8.south west);
\coordinate (kb) at ($(K0.north west)!0.5!(K8.south west)+(-5.5pt,0)$);
\draw[dflow] (B3.east) to[out=0,in=180] (kb);

\node[anchor=south, inner sep=0pt, font=\tiny] at (37.8,-88.5)
  {\cv{$\mathsf{conv}(x)[n,\cdot,\cdot,w]$}};
\foreach \h/\amin/\vals in {%
  0/10/{-0.01,+0.12,+0.06,-0.01,+0.18,-0.10,-0.10,-0.08,+0.06,+0.06,-0.11,+0.04,+0.13,+0.13,+0.00,+0.08},
  1/6/{+0.10,-0.04,-0.04,-0.04,+0.02,-0.03,-0.20,+0.03,-0.13,-0.05,+0.15,-0.17,+0.04,-0.11,-0.08,-0.02},
  2/11/{-0.04,+0.06,+0.01,-0.04,-0.08,-0.08,-0.02,+0.08,-0.03,-0.05,+0.12,-0.11,+0.03,-0.08,+0.04,+0.11},
  3/9/{-0.11,+0.03,+0.17,+0.26,+0.11,-0.05,+0.22,-0.07,+0.04,-0.15,+0.06,-0.10,+0.10,+0.02,+0.15,+0.12},
  4/8/{+0.18,-0.05,+0.06,+0.25,+0.05,-0.07,-0.16,+0.01,-0.22,+0.17,+0.07,-0.18,-0.02,-0.12,-0.02,+0.14},
  5/3/{+0.01,+0.30,+0.02,-0.16,+0.16,+0.15,+0.04,+0.22,-0.14,+0.17,-0.12,-0.01,+0.16,+0.06,-0.11,+0.31},
  6/14/{+0.03,+0.12,+0.05,+0.00,+0.11,+0.05,+0.02,+0.06,-0.07,-0.06,-0.05,-0.08,-0.08,+0.07,-0.08,+0.12},
  7/10/{-0.10,+0.32,-0.01,-0.16,+0.11,-0.15,-0.08,-0.08,+0.23,+0.20,-0.22,+0.13,-0.15,+0.14,+0.08,+0.16},
  8/10/{+0.05,+0.18,+0.10,+0.02,+0.37,-0.02,-0.08,-0.14,+0.06,+0.07,-0.28,+0.10,+0.08,+0.29,-0.09,+0.08},
  9/6/{-0.01,-0.06,-0.09,-0.21,-0.06,-0.10,-0.37,-0.08,-0.07,+0.04,+0.20,-0.15,-0.01,-0.18,-0.07,-0.10},
  10/5/{-0.13,+0.10,+0.01,-0.13,-0.05,-0.20,-0.08,-0.02,+0.13,-0.02,+0.08,+0.01,+0.21,-0.01,+0.12,+0.06},
  11/5/{+0.12,+0.04,-0.09,+0.17,+0.05,-0.21,+0.08,-0.04,-0.07,-0.13,+0.05,-0.12,-0.05,+0.01,-0.03,-0.02},
  12/5/{-0.02,+0.07,+0.04,+0.00,+0.05,-0.16,-0.13,-0.05,+0.04,+0.07,+0.01,-0.01,+0.16,+0.01,+0.06,+0.08},
  13/11/{-0.00,+0.09,-0.06,+0.05,+0.09,-0.08,+0.08,-0.03,-0.07,-0.05,+0.06,-0.12,+0.07,-0.03,-0.02,+0.05},
  14/5/{+0.02,+0.04,+0.05,+0.05,+0.05,-0.15,-0.14,-0.04,-0.01,+0.10,+0.03,-0.05,+0.14,-0.02,+0.05,+0.09},
  15/8/{+0.02,+0.13,-0.07,+0.00,+0.09,-0.06,+0.07,+0.02,-0.11,+0.00,+0.02,-0.10,+0.07,-0.01,-0.06,+0.09},
  16/5/{-0.01,+0.07,+0.04,+0.01,+0.06,-0.14,-0.11,-0.04,+0.02,+0.06,+0.01,-0.02,+0.13,+0.02,+0.05,+0.08},
  17/8/{+0.18,+0.04,-0.09,-0.01,+0.01,-0.08,-0.13,+0.10,-0.20,-0.01,+0.06,-0.14,+0.00,-0.03,-0.15,+0.04},
  18/3/{-0.10,+0.12,+0.01,-0.13,-0.04,-0.09,+0.00,+0.05,+0.07,-0.09,+0.06,-0.03,+0.08,+0.01,+0.05,+0.08},
  19/5/{+0.06,+0.09,+0.03,+0.13,+0.12,-0.12,+0.02,-0.03,+0.08,-0.08,-0.04,-0.04,-0.07,+0.06,+0.08,+0.06},
  20/8/{+0.11,+0.06,+0.06,+0.11,+0.14,+0.02,-0.05,+0.02,-0.13,+0.04,-0.05,-0.10,-0.06,+0.03,-0.08,+0.13},
  21/3/{-0.09,+0.14,+0.03,-0.11,+0.01,-0.07,-0.08,-0.01,-0.02,+0.13,-0.03,-0.02,+0.02,+0.01,-0.02,+0.15},
  22/5/{-0.04,+0.12,+0.05,-0.04,+0.12,-0.13,-0.10,-0.07,+0.08,+0.04,-0.06,+0.03,+0.16,+0.10,+0.04,+0.07},
  23/6/{+0.10,+0.11,-0.10,-0.06,+0.12,-0.06,-0.10,+0.02,-0.02,-0.01,+0.01,-0.08,-0.02,-0.02,-0.04,+0.02},
  24/11/{+0.02,+0.10,+0.04,+0.01,+0.08,-0.02,-0.03,+0.03,-0.04,-0.01,-0.02,-0.06,+0.00,+0.04,-0.03,+0.11},
  25/6/{+0.13,+0.07,+0.01,-0.06,+0.02,-0.05,-0.22,+0.10,-0.08,+0.05,-0.02,-0.06,-0.05,+0.01,-0.08,+0.10},
  26/9/{-0.05,+0.14,+0.03,-0.10,+0.03,+0.02,+0.05,+0.08,+0.00,-0.12,-0.00,-0.05,-0.04,+0.05,-0.04,+0.11},
  27/10/{-0.08,+0.16,+0.11,+0.04,+0.15,-0.08,-0.02,-0.08,+0.23,+0.00,-0.10,+0.04,-0.08,+0.09,+0.17,+0.11},
  28/8/{+0.18,+0.06,+0.09,+0.19,+0.23,+0.07,-0.06,+0.00,-0.18,+0.08,-0.12,-0.10,-0.11,+0.08,-0.14,+0.14},
  29/0/{-0.23,+0.10,+0.17,-0.13,+0.10,+0.14,-0.06,-0.01,+0.03,+0.09,+0.06,-0.05,+0.18,-0.07,+0.10,+0.19},
  30/8/{+0.12,+0.00,+0.05,+0.16,+0.05,-0.03,-0.09,+0.04,-0.17,+0.08,+0.05,-0.15,-0.04,-0.08,-0.04,+0.13},
  31/10/{+0.08,+0.30,+0.03,-0.17,+0.17,+0.13,-0.07,+0.22,-0.06,+0.17,-0.17,+0.03,+0.06,+0.10,-0.08,+0.29}}{%
  \foreach [count=\c from 0] \v in \vals {%
    \pgfmathtruncatemacro{\p}{min(90, 12 + abs(\v)*230)}%
    \ifdim \v pt<0pt \colorlet{cellc}{cF!\p}\else \colorlet{cellc}{cA!\p}\fi
    \ifnum\c=\amin \colorlet{cellc}{cE!85}\fi
    \fill[cellc] ($(6,-90)+(\c*4,-\h*4)$) rectangle ++(3.6,-3.4);}
  \fill[cE!85] ($(84,-90)+(0,-\h*4)$) rectangle ++(5,-3.4);}
\draw[dflow, line width=0.4pt, -{Latex[length=1mm]}] (71,-154) -- (82,-154);
\node[anchor=south, inner sep=0pt, font=\tiny] at (76.5,-152) {\mn{min}};
\colorlet{resopt}{cE!85!black}
\colorlet{resref}{cE!55!black}
\draw[resopt, line width=0.7pt, -{Latex[length=1mm]}, rounded corners=2pt]
  (94,-90) -- (96,-90) -- (96,-153.4) -- (130,-153.4);
\node[anchor=south west, inner sep=1pt, font=\tiny] at (98.5,-152.9) {opt: \sm{sum} $h{<}32$};
\fill[resopt] (131,-151) rectangle ++(5,-5);
\node[anchor=west, inner sep=0pt, font=\tiny, text=resopt] (Nopt) at (138,-153.4) {$\mathbf{-5.1}$};
\draw[resref, line width=0.7pt, -{Latex[length=1mm]}, rounded corners=2pt]
  (90,-90) -- (92,-90) -- (92,-217.4) -- (130,-217.4);
\node[anchor=south west, inner sep=1pt, font=\tiny] at (94.5,-216.9) {ref: \sm{sum} $h{<}64$};
\fill[resref] (131,-214.9) rectangle ++(5,-5);
\node[anchor=west, inner sep=0pt, font=\tiny, text=resref] (Nref) at (138,-217.4) {$\mathbf{-10.5}$};

\begin{scope}[shift={(291.8,-195)}, x=8.6pt, y=24pt]
  \fill[cF!9]  (-13,-0.28) rectangle (-2.68,2.1);
  \fill[cB!16] (-2.68,-0.28) rectangle (2,2.1);
  \draw[cB!60, line width=0.4pt, dash pattern=on 1pt off 1pt] (-2.68,-0.28) -- (-2.68,2.1);
  \draw[cF, line width=0.4pt] (-13,0) -- (2.1,0);
  \foreach \s in {-8,-4,0}{%
    \draw[cF, line width=0.4pt] (\s,0) -- (\s,-0.04);
    \node[anchor=north, inner sep=0.5pt, font=\tiny, text=cF] at (\s,-0.05) {$\s$};}
  \draw[black, line width=0.6pt] plot coordinates {
    (-13.00,-0.0000) (-8.00,-0.0000) (-5.00,-0.0000) (-4.00,-0.0001) (-3.75,-0.0003)
    (-3.50,-0.0008) (-3.25,-0.0019) (-3.00,-0.0040) (-2.75,-0.0082) (-2.50,-0.0155)
    (-2.25,-0.0275) (-2.00,-0.0455) (-1.75,-0.0701) (-1.50,-0.1002) (-1.25,-0.1321)
    (-1.00,-0.1587) (-0.75,-0.1700) (-0.50,-0.1543) (-0.25,-0.1003) (0.00,0.0000)
    (0.25,0.1497) (0.50,0.3457) (0.75,0.5800) (1.00,0.8413) (1.25,1.1179)
    (1.50,1.3998) (1.75,1.6799) (2.00,1.9545)};
  \node[anchor=north west, inner sep=2pt, font=\tiny] at (-13,2.1) {\gl{$\mathsf{gelu}(s)$}};
  \node[anchor=north, inner sep=0pt, font=\tiny, text=cB, align=center] at (-0.6,1.3)
    {difference\\visible};
  \coordinate (Popt) at (-5.1,0);
  \coordinate (Pref) at (-10.5,0);
\end{scope}
\fill[resopt] (Popt) circle (1.2pt);
\fill[resref] (Pref) circle (1.2pt);
\draw[dflow, line width=0.5pt, -{Latex[length=1mm]}, resopt, shorten >=1.5pt] (Nopt.east) -- (Popt);
\draw[dflow, line width=0.5pt, -{Latex[length=1mm]}, resref, shorten >=1.5pt] (Nref.east) -- (Pref);
\end{tikzpicture}
\caption{A worked example of a real bug (KernelBench level-2 task 36, \archive row 5354).
\textbf{Above:} The reference program is optimized by fusing two operations into a custom kernel, but the kernel has an incorrect loop bound.
\textbf{Below:} Random inputs are highly unlikely to expose the bug because the \texttt{min} and \texttt{sum} operations conspire to produce values that are in the flat region of \GELU, masking the bug.
}
\label{fig:example}
\end{figure*}

%% file: fig-overview.tex
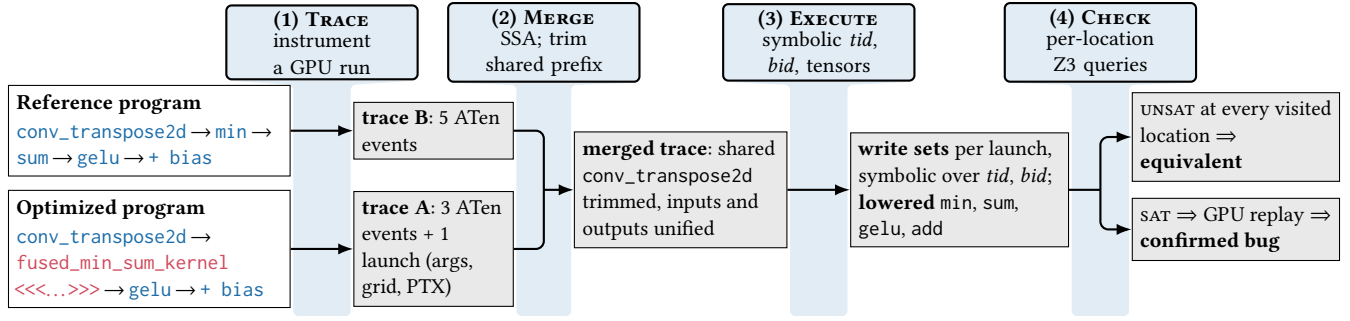
\begin{figure*}[t]
\centering
\def\ovarr{\,$\to$\,}
\begin{tikzpicture}[
  x=1pt, y=1pt,
  stg/.style={stage, text width=68pt, minimum height=0pt, inner sep=2pt, fill=zonec},
  zone/.style={fill=zonec, draw=none, rounded corners=2pt},
  art/.style={artifact, align=flush left, anchor=center},
  inp/.style={draw, fill=white, align=flush left, inner sep=3pt, text width=100pt,
              font=\footnotesize, anchor=center},
  leg/.style={thick, rounded corners=3pt},
  flowr/.style={flow, rounded corners=3pt},
]
\coordinate (x1) at (0,0);
\coordinate (x2) at (107,0);
\coordinate (x3) at (200,0);
\coordinate (x4) at (305,0);
\coordinate (x5) at (409,0);

\node[inp, anchor=north] (ref) at (x1) {%
  \textbf{Reference program}\\[1pt]
  \ttfamily
  \textcolor{cA}{conv\_transpose2d}\ovarr\textcolor{cA}{min}\ovarr\\
  \textcolor{cA}{sum}\ovarr\textcolor{cA}{gelu}\ovarr\textcolor{cA}{+ bias}};
\node[inp, below=5pt of ref] (opt) {%
  \textbf{Optimized program}\\[1pt]
  \ttfamily
  \textcolor{cA}{conv\_transpose2d}\ovarr\\
  \textcolor{cB}{fused\_min\_sum\_kernel}\\
  \textcolor{cB}{<{}<{}<{\rmfamily\ldots}>{}>{}>}\ovarr\textcolor{cA}{gelu}\ovarr\textcolor{cA}{+ bias}};
\coordinate (mid) at ($(ref.center)!0.5!(opt.center)$);

\node[art, text width=54pt] (trA) at (x2 |- ref) {\textbf{trace B}:
  5 ATen events};
\node[art, text width=54pt] (trB) at (x2 |- opt) {\textbf{trace A}:
  3 ATen events + 1 launch (args, grid, PTX)};

\node[art, text width=74pt] (merged) at (x3 |- mid) {\textbf{merged trace}:
  shared \texttt{conv\_transpose2d} trimmed, inputs and outputs unified};
\node[art, text width=76pt] (ws) at (x4 |- mid) {\textbf{write sets} per launch,
  symbolic over \tid, \bid; \textbf{lowered} \texttt{min}, \texttt{sum}, \texttt{gelu}, \texttt{add}};

\coordinate (upper) at ($(mid)+(0,3)$);
\coordinate (lower) at ($(mid)-(0,3)$);
\node[art, text width=72pt, anchor=south] (out1) at (x5 |- upper) {\unsat{} at every visited
  location $\Rightarrow$ \textbf{equivalent}};
\node[art, text width=72pt, anchor=north] (out2) at (x5 |- lower) {\sat{} $\Rightarrow$ GPU replay
  $\Rightarrow$ \textbf{confirmed bug}};

\coordinate (m1) at ($(ref.east)!0.5!(trA.west)$);
\coordinate (g2) at ($(trA.east)!0.5!(merged.west)$);
\coordinate (j2) at (g2 |- mid);
\coordinate (m3) at ($(merged.east)!0.5!(ws.west)$);
\coordinate (g4) at ($(ws.east)!0.5!(out1.west)$);
\coordinate (j4) at (g4 |- mid);

\coordinate (hdr) at (0,16);
\node[stg] (tracer)   at (m1 |- hdr) {\textbf{(1) \TRACE}\\[-2pt]\footnotesize instrument a GPU run};
\node[stg] (merger)   at (g2 |- hdr) {\textbf{(2) \MERGE}\\[-2pt]\footnotesize SSA; trim shared prefix};
\node[stg] (executor) at (m3 |- hdr) {\textbf{(3) \EXECUTE}\\[-2pt]\footnotesize symbolic \tid, \bid, tensors};
\node[stg] (checker)  at (g4 |- hdr) {\textbf{(4) \CHECK}\\[-2pt]\footnotesize per-location Z3 queries};

\coordinate (lanetop) at (0,0);
\coordinate (lanebot) at ($(opt.south)+(0,-3)$);
\coordinate (l1) at ($(ref.east)+(1,0)$);    \coordinate (r1) at ($(trA.west)-(1,0)$);
\coordinate (l2) at ($(trA.east)+(1,0)$);    \coordinate (r2) at ($(merged.west)-(1,0)$);
\coordinate (l3) at ($(merged.east)+(1,0)$); \coordinate (r3) at ($(ws.west)-(1,0)$);
\coordinate (l4) at ($(ws.east)+(1,0)$);     \coordinate (r4) at ($(out1.west)-(1,0)$);
\begin{scope}[on background layer]
  \fill[zone] (tracer.south west) -- (tracer.south east)
    -- (r1 |- lanetop) -- (r1 |- lanebot) -- (l1 |- lanebot) -- (l1 |- lanetop) -- cycle;
  \fill[zone] (merger.south west) -- (merger.south east)
    -- (r2 |- lanetop) -- (r2 |- lanebot) -- (l2 |- lanebot) -- (l2 |- lanetop) -- cycle;
  \fill[zone] (executor.south west) -- (executor.south east)
    -- (r3 |- lanetop) -- (r3 |- lanebot) -- (l3 |- lanebot) -- (l3 |- lanetop) -- cycle;
  \fill[zone] (checker.south west) -- (checker.south east)
    -- (r4 |- lanetop) -- (r4 |- lanebot) -- (l4 |- lanebot) -- (l4 |- lanetop) -- cycle;
\end{scope}

\draw[flow] (ref.east) -- (trA.west);
\draw[flow] (opt.east) -- (trB.west);

\draw[leg] (trA.east) -- (j2 |- trA.east) -- (j2);
\draw[leg] (trB.east) -- (j2 |- trB.east) -- (j2);
\draw[flow] (j2) -- (merged.west);

\draw[flow] (merged.east) -- (ws.west);

\draw[leg]   (ws.east) -- (j4);
\draw[flowr] (j4) -- (j4 |- out1.west) -- (out1.west);
\draw[flowr] (j4) -- (j4 |- out2.west) -- (out2.west);

\end{tikzpicture}
\caption{
  The running example's path through \dirigo.
  (1) Each program is run on the GPU to extract its trace.
  (2) The two traces are merged into one equivalence-checking input.
  (3) Each kernel launch is lowered to symbolic expressions by a scalable PTX-level symbolic execution.
  (4) The expressions are specialized to individual output locations and checked for equivalence over all inputs.}
\label{fig:overview}
\end{figure*}

%% file: fig-trace-merge.tex
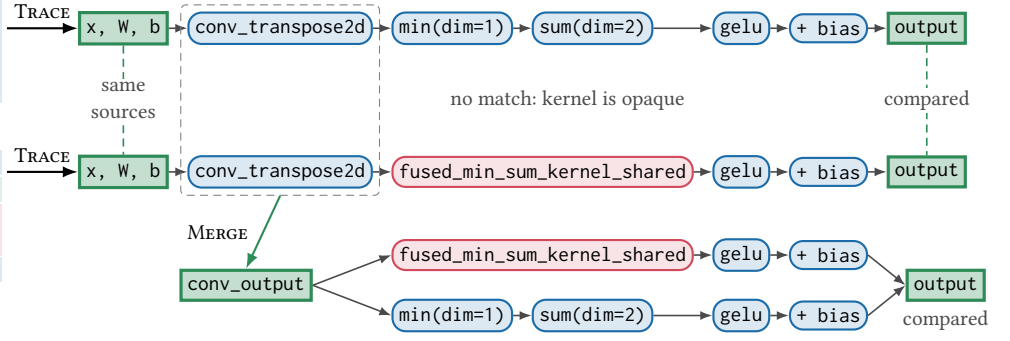
\begin{figure*}[t]
\centering
\begin{tikzpicture}[
  code/.style={anchor=north west, outer sep=0pt, inner xsep=4pt, inner ysep=1.2pt,
               font=\footnotesize\ttfamily, align=left, text width=124pt},
  bar/.style={path picture={\fill[#1] (path picture bounding box.north west)
              rectangle ([xshift=2.2pt]path picture bounding box.south west);}},
  op/.style={torchop, inner xsep=2.5pt, inner ysep=2.6pt},
  ko/.style={cudaop, inner xsep=2.5pt, inner ysep=2.6pt},
  io2/.style={iotensor, font=\footnotesize\ttfamily, inner xsep=2.5pt, inner ysep=2.6pt, align=center},
  corr/.style={densely dashed, cF!70, semithick},
  tlab/.style={font=\footnotesize, text=cF, align=center, inner sep=1pt, fill=white},
  hdr/.style={anchor=south west, inner sep=0pt, font=\footnotesize\bfseries},
  node distance=1pt,
]
\node[code, fill=cAl, bar=cA, anchor=west] (ref) at (0,0)
  {x = conv\_transpose2d(x, W, b)\\
   x = min(x, dim=1)\\
   x = sum(x, dim=2)\\
   x = gelu(x) + bias};
\node[hdr] at ([yshift=2pt]ref.north west) {Reference program (Python)};

\node[code, fill=cAl, bar=cA] (o1) at (0,-1.32)
  {x = conv\_transpose2d(x, W, b);};
\node[code, fill=cCl, bar=cC, below=of o1] (o2)
  {out = torch::zeros(\{N,1,1,W\});};
\node[code, fill=cBl, bar=cB, below=of o2] (o3)
  {fused\_min\_sum\_kernel\_shared\\
   \hspace*{1em}<{}<{}<(N,W), 256>{}>{}>(x, out);};
\node[code, fill=cAl, bar=cA, below=of o3] (o4)
  {out = gelu(out) + bias;};
\node[hdr] at ([yshift=2pt]o1.north west) {Optimized program (C++ host)};

\node[io2] (b-in) at (6.25,0.3) {x, W, b};
\node[op, right=7pt of b-in]   (b-conv) {conv\_transpose2d};
\node[op, right=7pt of b-conv] (b-min)  {min(dim=1)};
\node[op, right=7pt of b-min]  (b-sum)  {sum(dim=2)};
\node[io2] (a-in)  at ($(b-in)+(0,-1.9)$)  {x, W, b};
\node[op]  (a-conv) at ($(b-conv)+(0,-1.9)$) {conv\_transpose2d};
\node[ko, right=7pt of a-conv] (a-fus) {fused\_min\_sum\_kernel\_shared};
\path let \p1 = (b-sum.east), \p2 = (a-fus.east) in
  node[op, anchor=west] (b-gelu) at ({max(\x1,\x2)+7pt}, \y1) {gelu};
\node[op, right=7pt of b-gelu] (b-add)  {+ bias};
\node[io2, right=7pt of b-add] (b-out)  {output};
\draw[dflow] (b-in) -- (b-conv);  \draw[dflow] (b-conv) -- (b-min);
\draw[dflow] (b-min) -- (b-sum);  \draw[dflow] (b-sum) -- (b-gelu);
\draw[dflow] (b-gelu) -- (b-add); \draw[dflow] (b-add) -- (b-out);
\draw[flow] ([xshift=2pt]ref.east |- b-in) -- node[above, font=\footnotesize] {\TRACE} (b-in.west);

\node[op]  (a-gelu) at ($(b-gelu)+(0,-1.9)$) {gelu};
\node[op]  (a-add) at ($(b-add)+(0,-1.9)$)  {+ bias};
\node[io2] (a-out) at ($(b-out)+(0,-1.9)$)  {output};
\draw[dflow] (a-in) -- (a-conv);  \draw[dflow] (a-conv) -- (a-fus);
\draw[dflow] (a-fus) -- (a-gelu); \draw[dflow] (a-gelu) -- (a-add);
\draw[dflow] (a-add) -- (a-out);
\draw[flow] ([xshift=2pt]o2.east |- a-in) -- node[above, font=\footnotesize] {\TRACE} (a-in.west);

\draw[corr, cC] (b-in.south) -- node[tlab] {same\\sources} (a-in.north);
\node[draw=cF!70, densely dashed, rounded corners=2pt, inner sep=2.5pt,
      fit=(b-conv)(a-conv)] (cbox) {};
\node[tlab] at ($(b-sum.south)!0.5!(a-fus.north)$)
  {no match: kernel is opaque};
\draw[corr, cC] (b-out.south) -- node[tlab] {compared} (a-out.north);

\node[io2] (m-mp)  at ($(a-conv)+(-0.45,-1.5)$) {conv\_output};
\node[ko]  (m-fus) at ($(a-fus)+(0,-1.1)$) {fused\_min\_sum\_kernel\_shared};
\node[op]  (mu-gelu) at (b-gelu |- m-fus) {gelu};
\node[op]  (mu-add)  at (b-add |- m-fus)  {+ bias};
\node[op]  (ml-min)  at ($(b-min)+(0,-3.8)$) {min(dim=1)};
\node[op]  (ml-sum)  at (b-sum |- ml-min) {sum(dim=2)};
\node[op]  (ml-gelu) at (b-gelu |- ml-min) {gelu};
\node[op]  (ml-add)  at (b-add |- ml-min)  {+ bias};
\node[io2] (m-out) at ($(b-out)+(0.25,-3.4)$) {output};
\draw[dflow] (m-mp.east) -- (m-fus.west);
\draw[dflow] (m-mp.east) -- (ml-min.west);
\draw[dflow] (m-fus) -- (mu-gelu);  \draw[dflow] (mu-gelu) -- (mu-add);
\draw[dflow] (ml-min) -- (ml-sum);  \draw[dflow] (ml-sum) -- (ml-gelu);
\draw[dflow] (ml-gelu) -- (ml-add);
\draw[dflow] (mu-add.east) -- (m-out.west);
\draw[dflow] (ml-add.east) -- (m-out.west);
\node[tlab, anchor=north] at ([yshift=-2pt]m-out.south) {compared};

\draw[flow, cC] (cbox.south) -- node[left=2pt, font=\footnotesize, text=black] {\MERGE}
  (m-mp.north);

\node[op, minimum width=11pt, minimum height=8pt, anchor=west] (lgA) at (0,-3.32) {};
\node[anchor=west, font=\footnotesize, inner xsep=3pt] at (lgA.east) {ATen op};
\node[iotensor, minimum width=11pt, minimum height=8pt, anchor=west] (lgC) at (2.3,-3.32) {};
\node[anchor=west, font=\footnotesize, inner xsep=3pt] at (lgC.east) {tensor};
\node[ko, minimum width=11pt, minimum height=8pt, anchor=west] (lgB) at (0,-3.7) {};
\node[anchor=west, font=\footnotesize, inner xsep=3pt] at (lgB.east) {CUDA kernel launch};
\end{tikzpicture}
\caption{
\TRACE and \MERGE on the running example.
\textbf{Left:} the reference and optimized programs.
\textbf{Top right:} \TRACE records each program's Torch dispatches and CUDA events via \texttt{TorchDispatchMode}, \texttt{at::RecordFunction}, and CUDA-runtime \texttt{LD\_PRELOAD}.
\textbf{Bottom:} \MERGE trims pure prefix operations when they're common between both programs.}
\label{fig:trace-merge}
\end{figure*}

%% file: fig-executor.tex
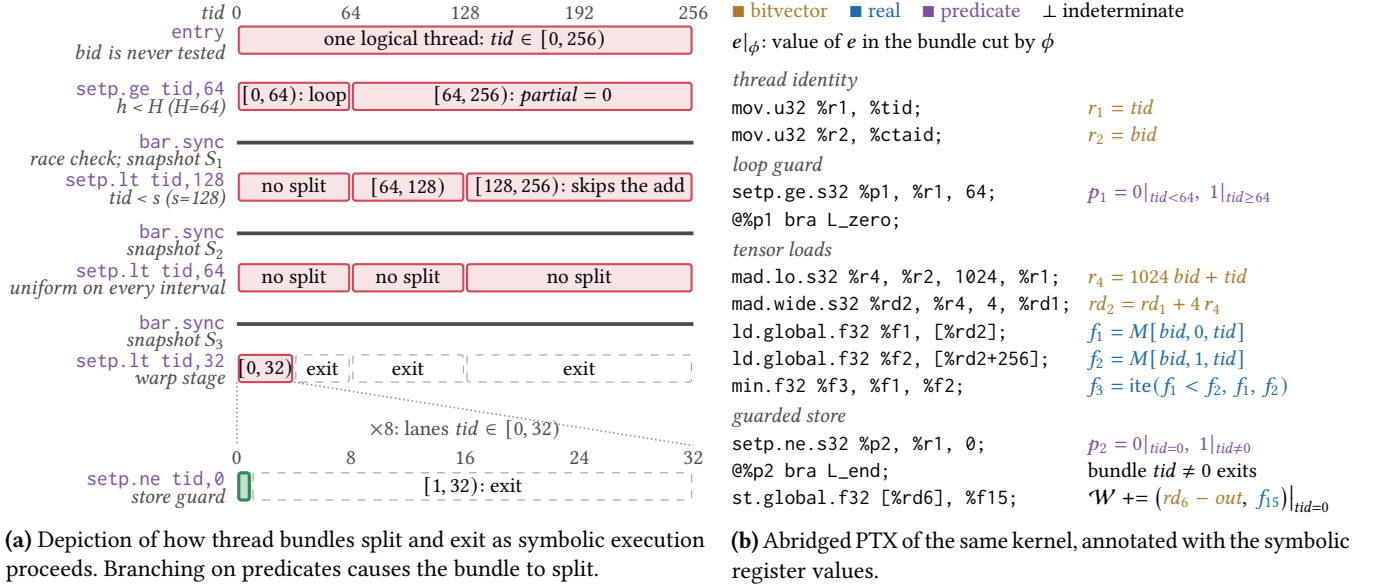
\begin{figure*}[t]
\centering
\begin{subfigure}[t]{0.52\textwidth}
\centering
\begin{tikzpicture}[x=1cm, y=1.2cm, baseline=(current bounding box.north),
  live/.style={draw=cB, thick, fill=cBl, rounded corners=1pt},
  gone/.style={draw=cF!55, dashed, fill=white, rounded corners=1pt},
  leaf/.style={draw=cC, very thick, fill=cC!28, rounded corners=1pt},
  seglab/.style={font=\footnotesize, inner sep=0pt},
  rowlab/.style={anchor=east, font=\footnotesize\ttfamily, inner sep=1pt, align=right, text=cE},
  rowsub/.style={anchor=east, font=\footnotesize\itshape, inner sep=1pt, align=right, text=cF},
  tick/.style={font=\footnotesize, text=cF, inner sep=1pt},
  barrier/.style={cF, line width=1.4pt},
  zoom/.style={cF!60, densely dotted, semithick},
]
\def\s{0.023594}\def\z{0.18875}
\newcommand{\seg}[6]{%
  \draw[#1] ([xshift=0.7pt]{#3*#5},#2) rectangle ([xshift=-0.7pt]{#4*#5},#2+0.3);
  \node[seglab] at ({(#3+#4)/2*#5},#2+0.15) {#6};}
\newcommand{\brow}[3]{%
  \draw[barrier] (0,#1) -- (6.04,#1);
  \node[rowlab] at (-0.12,#1) {#2};
  \node[rowsub] at (-0.12,#1-0.19) {#3};}
\newcommand{\rowl}[3]{%
  \node[rowlab] at (-0.12,#1+0.22) {#2};
  \node[rowsub] at (-0.12,#1+0.02) {#3};}
\foreach \t in {0,64,128,192,256}
  \node[tick, anchor=south] at ({\t*\s},0.34) {\t};
\node[tick, anchor=south east] at (-0.12,0.34) {\tid};
\rowl{0}{entry}{\bid is never tested}
\seg{live}{0}{0}{256}{\s}{one logical thread: $\tid\in[0,256)$}
\rowl{-0.62}{setp.ge tid,64}{h < H\ (H=64)}
\seg{live}{-0.62}{0}{64}{\s}{$[0,64)$: loop}
\seg{live}{-0.62}{64}{256}{\s}{$[64,256)$: $\mathit{partial}=0$}
\brow{-0.98}{bar.sync}{race check; snapshot $S_1$}
\rowl{-1.62}{setp.lt tid,128}{tid < s\ (s=128)}
\seg{live}{-1.62}{0}{64}{\s}{no split}
\seg{live}{-1.62}{64}{128}{\s}{$[64,128)$}
\seg{live}{-1.62}{128}{256}{\s}{$[128,256)$: skips the add}
\brow{-1.98}{bar.sync}{snapshot $S_2$}
\rowl{-2.62}{setp.lt tid,64}{uniform on every interval}
\seg{live}{-2.62}{0}{64}{\s}{no split}
\seg{live}{-2.62}{64}{128}{\s}{no split}
\seg{live}{-2.62}{128}{256}{\s}{no split}
\brow{-2.98}{bar.sync}{snapshot $S_3$}
\rowl{-3.62}{setp.lt tid,32}{warp stage}
\seg{live}{-3.62}{0}{32}{\s}{$[0,32)$}
\seg{gone}{-3.62}{32}{64}{\s}{exit}
\seg{gone}{-3.62}{64}{128}{\s}{exit}
\seg{gone}{-3.62}{128}{256}{\s}{exit}
\draw[zoom] (0,-3.62) -- (0,-4.32);
\draw[zoom] ({32*\s},-3.62) -- (6.04,-4.32);
\node[tick, anchor=south] at (3.02,-4.28) {$\times 8$: lanes $\tid\in[0,32)$};
\foreach \t in {0,8,16,24,32}
  \node[tick, anchor=north] at ({\t*\z},-4.34) {\t};
\rowl{-4.92}{setp.ne tid,0}{store guard}
\seg{leaf}{-4.92}{0}{1}{\z}{}
\seg{gone}{-4.92}{1}{32}{\z}{$[1,32)$: exit}
\end{tikzpicture}
\caption{Depiction of how thread bundles split and exit as symbolic execution proceeds. Branching on predicates causes the bundle to split.}
\label{fig:bundles-a}
\end{subfigure}\hfill
\begin{subfigure}[t]{0.46\textwidth}
\newcommand{\bv}[1]{{\color{cD!75!black}$#1$}}
\newcommand{\re}[1]{{\color{cA}$#1$}}
\newcommand{\pr}[1]{{\color{cE}$#1$}}
\newcommand{\ins}[1]{{\ttfamily #1}}
\newcommand{\hd}[1]{\multicolumn{2}{@{}l@{}}{\rule{0pt}{8.5pt}{\itshape\color{cF}#1}}\\}
{\footnotesize\color{cD!75!black}$\blacksquare$ bitvector\quad
 \color{cA}$\blacksquare$ real\quad
 \color{cE}$\blacksquare$ predicate\quad
 \color{black}$\botval$ indeterminate\\
 $e|_{\phi}$: value of $e$ in the bundle cut by $\phi$}\par
\smallskip
\renewcommand{\arraystretch}{1.02}
{\footnotesize
\begin{tabular}{@{}l@{\hspace{6pt}}l@{}}
\hd{thread identity}
\ins{mov.u32 \%r1, \%tid;}                & \bv{r_1=\tid}\\
\ins{mov.u32 \%r2, \%ctaid;}              & \bv{r_2=\bid}\\
\hd{loop guard}
\ins{setp.ge.s32 \%p1, \%r1, 64;}         & \pr{p_1=0|_{\tid<64},\ 1|_{\tid\ge64}}\\
\ins{@\%p1 bra L\_zero;}                  & \\
\hd{tensor loads}
\ins{mad.lo.s32 \%r4, \%r2, 1024, \%r1;}  & \bv{r_4=1024\,\bid+\tid}\\
\ins{mad.wide.s32 \%rd2, \%r4, 4, \%rd1;} & \bv{\mathit{rd}_2=\mathit{rd}_1+4\,r_4}\\
\ins{ld.global.f32 \%f1, [\%rd2];}        & \re{f_1=M[\bid,0,\tid]}\\
\ins{ld.global.f32 \%f2, [\%rd2+256];}    & \re{f_2=M[\bid,1,\tid]}\\
\ins{min.f32 \%f3, \%f1, \%f2;}           & \re{f_3=\ite{f_1<f_2}{f_1}{f_2}}\\
\hd{guarded store}
\ins{setp.ne.s32 \%p2, \%r1, 0;}          & \pr{p_2=0|_{\tid=0},\ 1|_{\tid\ne0}}\\
\ins{@\%p2 bra L\_end;}                   & bundle $\tid\ne0$ exits\\
\ins{st.global.f32 [\%rd6], \%f15;}       & $\wset \mathrel{+}= \big({\color{cD!75!black}\mathit{rd}_6-\mathit{out}},\ {\color{cA}f_{15}}\big)\big|_{\tid=0}$\\
\end{tabular}}
\caption{Abridged PTX of the same kernel, annotated with the symbolic register values.}
\label{fig:bundles-b}
\end{subfigure}
\caption{Running \EXECUTE on the kernel in the running example.}
\label{fig:bundles}
\end{figure*}

%% file: fig-checker.tex
\providecommand{\hlb}[1]{{\setlength{\fboxsep}{0.6pt}\colorbox{cB!25}{$\scriptstyle\mathbf{#1}$}}}
\begin{figure*}[t]
\centering
\def\gs{0.21}%
\begin{tikzpicture}[
  x=1cm, y=1cm,
  panel/.style={draw=cF!45, rounded corners=3pt, inner sep=4pt},
  colhead/.style={font=\footnotesize\itshape, text=cF, inner sep=1pt},
  rev/.style={-{Latex[length=1.5mm]}, semithick, cF},
  link/.style={-{Latex[length=1.5mm]}, semithick, cB, dashed},
  q/.style={font=\footnotesize, fill=white, inner sep=1.5pt, midway},
  top/.style={torchop, font=\footnotesize, align=flush left, text width=5.7cm},
  kop/.style={cudaop, font=\footnotesize, align=flush left, text width=5.7cm},
  cmp/.style={draw=cD, fill=cDl, rounded corners=2pt, align=flush left, font=\footnotesize, inner sep=4pt, minimum width=7.6cm},
]
\begin{scope}[shift={(0.5,-0.62)}]
  \coordinate (g0) at (0.05,0.3);
  \coordinate (g1) at (16*\gs,-32*\gs-0.05);
  \foreach \r in {0,...,31} \foreach \c in {0,...,15}
    \draw[cF!30, line width=0.25pt] (\c*\gs,-\r*\gs) rectangle ++(\gs,-\gs);
  \foreach \r/\c in {5/13,6/6,7/7,7/12,9/3,10/14,13/1,14/9,15/2,18/13,19/5,20/4,21/8,23/1,23/3,25/1,26/2,27/1,29/6}
    \fill[cC] (\c*\gs+0.031,-\r*\gs-0.031) rectangle ++(0.148,-0.148);
  \fill[cB] (3*\gs+0.031,-3*\gs-0.031) rectangle ++(0.148,-0.148);
  \draw[cB, very thick] (3*\gs,-3*\gs) rectangle ++(\gs,-\gs);
  \coordinate (rk) at (4*\gs,-3.5*\gs);
  \node[font=\footnotesize, anchor=south, inner sep=1pt] at (8*\gs,0.02) {$w$ (64 columns)};
  \node[font=\footnotesize, anchor=south, inner sep=1pt, rotate=90] at (-0.04,-16*\gs) {$n$ (128 rows)};
\end{scope}
\node[colhead] (ha) at (7.45,-0.22) {program A: optimized};
\node[colhead] (hb) at (13.86,-0.22) {program B: reference};
\node[iotensor, font=\footnotesize] (a0) at (7.45,-0.75) {$\ell = Y[11,5,12]$};
\node[top] (a1) at (7.45,-1.55) {\textbf{\ttfamily aten::add}~~$Y[n,c,w] = G[n,w] + b[c]$};
\node[top] (a2) at (7.45,-2.57) {\textbf{\ttfamily aten::gelu}~~$G[n,w] = \mathsf{gelu}(S[n,w])$};
\node[kop] (a3) at (7.45,-4.10) {\textbf{\ttfamily fused\_min\_sum\_kernel}\\
  $S[\bid_x,\bid_y] = \sum_{h<32}\min_{c'} M[\bid_x,c',h,\bid_y]$};
\node[iotensor, font=\footnotesize] (a4) at (7.45,-5.63) {$M$ (source tensor)};
\draw[rev] (a0) -- (a1);
\draw[rev] (a1) -- node[q] {$G[11,12] + b[5]$} (a2);
\draw[rev] (a2) -- node[q] {$\mathsf{gelu}(S[11,12]) + b[5]$} (a3);
\draw[rev] (a3) -- node[q] {reads $M[11,c',h,12]$, $h<32$} (a4);
\node[iotensor, font=\footnotesize] (b0) at (13.86,-0.75) {$\ell = Y[11,5,12]$};
\node[top] (b1) at (13.86,-1.55) {\textbf{\ttfamily aten::add}~~$Y[n,c,w] = G[n,w] + b[c]$};
\node[top] (b2) at (13.86,-2.57) {\textbf{\ttfamily aten::gelu}~~$G[n,w] = \mathsf{gelu}(S[n,w])$};
\node[top] (b3) at (13.86,-3.59) {\textbf{\ttfamily aten::sum}~~$S[n,w] = \sum_{h<64} R[n,h,w]$};
\node[top] (b4) at (13.86,-4.61) {\textbf{\ttfamily aten::min}~~$R[n,h,w] = \min_{c'} M[n,c',h,w]$};
\node[iotensor, font=\footnotesize] (b5) at (13.86,-5.63) {$M$ (source tensor)};
\draw[rev] (b0) -- (b1);
\draw[rev] (b1) -- node[q] {$G[11,12] + b[5]$} (b2);
\draw[rev] (b2) -- node[q] {$\mathsf{gelu}(S[11,12]) + b[5]$} (b3);
\draw[rev] (b3) -- node[q] {$\mathsf{gelu}\big(\sum_{h<64} R[11,h,12]\big) + b[5]$} (b4);
\draw[rev] (b4) -- node[q] {reads $M[11,c',h,12]$, $h<64$} (b5);
\node[cmp, anchor=north] (ex) at (10.655,-6.28) {%
  $e_A = \mathsf{gelu}\big(\sum_{h<\hlb{32}} \min_{c'} M[11,c',h,12]\big) + b[5]$\\[2pt]
  $e_B = \mathsf{gelu}\big(\sum_{h<\hlb{64}} \min_{c'} M[11,c',h,12]\big) + b[5]$\\[3pt]
  \textbf{check} $e_A = e_B$ for all $M$, $b$};
\draw[rev] (a4.south) |- ([yshift=3pt]ex.north -| a4);
\draw[rev] (b5.south) |- ([yshift=3pt]ex.north -| b5);
\draw[link] (rk.east) to[out=15,in=180] (a0.west);
\coordinate (pa1) at (0,-0.15);      \coordinate (pa2) at (3.90,0 |- ex.south);
\coordinate (pb1) at (4.48,-0.15);   \coordinate (pb2) at (16.90,0 |- ex.south);
\begin{scope}[on background layer]
  \node[panel, fit=(pa1)(pa2)(g0)(g1)(rk)] {};
  \node[panel, fit=(pb1)(pb2)(ha)(hb)(a0)(a3)(b1)(b5)(ex)] {};
\end{scope}
\end{tikzpicture}
\caption{Illustration of the \CHECK phase generating equivalence conditions for the running example. A heuristic searches concrete output locations, here $\ell = Y[11,5,12]$ (left), and walks both programs backwards from $\ell$ using \EXECUTE results to specialize each block's symbolic equation to the chosen location (right). An SMT solver searches for counterexamples to equivalence of the two resulting expressions.}
\label{fig:checker}
\end{figure*}
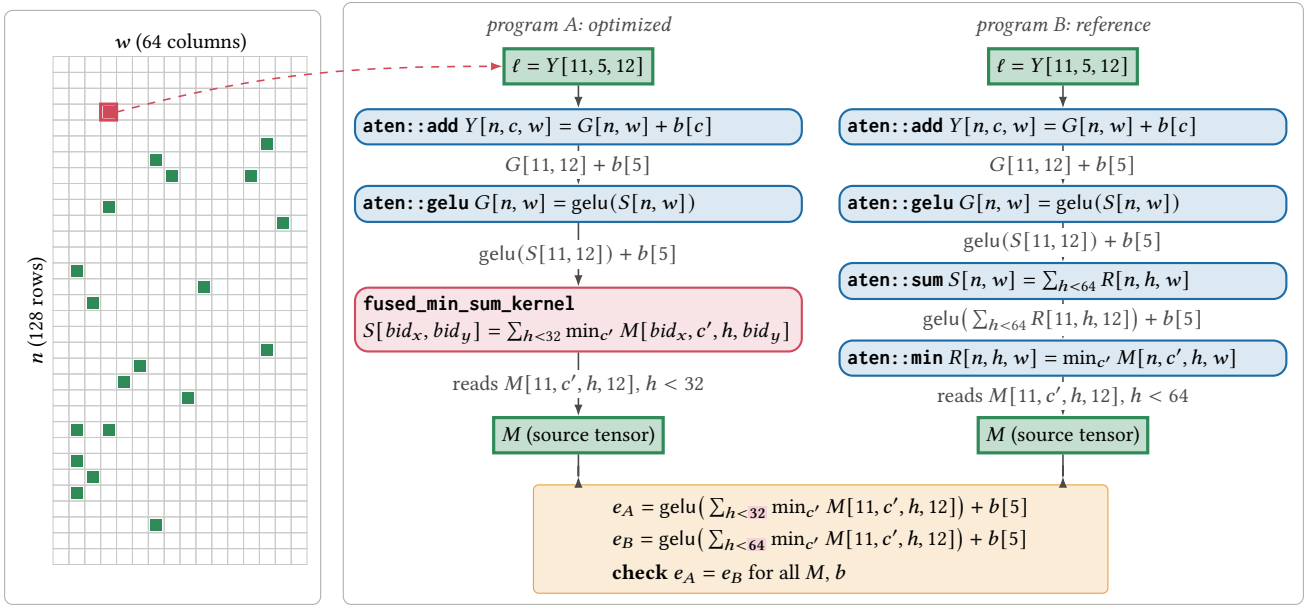

%% file: fig-corpus.tex
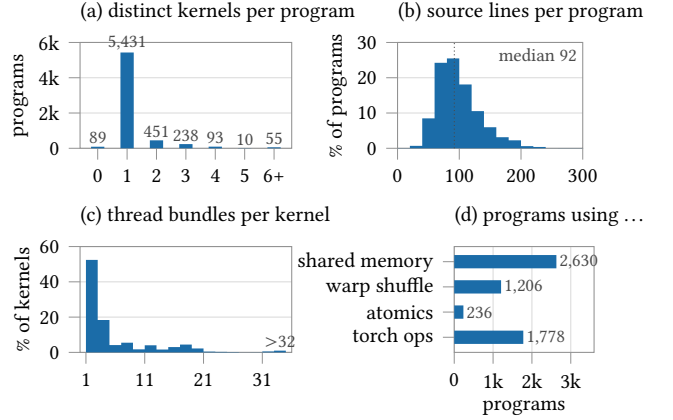
\begin{figure}[t]
  \centering
  \begin{tikzpicture}[
      panel title/.style={font=\footnotesize, anchor=south west, inner sep=0pt, yshift=1.5pt},
      count label/.style={font=\scriptsize, inner sep=1pt, text=cF},
      corner label/.style={count label, anchor=north east, inner sep=2pt},
    ]
    \pgfplotsset{
      corpus panel/.style={/pgfplots/.cd,
        scale only axis, width=2.5cm, height=1.4cm,
        label style={font=\fontsize{8}{9}\selectfont}, tick label style={font=\fontsize{8}{9}\selectfont},
        xlabel style={yshift=2pt}, ylabel style={yshift=-3pt},
        xtick pos=left, ytick pos=left,
        xtick align=outside, ytick align=outside,
        xtick style={draw=cF!60}, ytick style={draw=cF!60},
        axis line style={cF!60},
        ymajorgrids, grid style={cF!20},
        clip=false, unbounded coords=discard,
        title style={panel title, at={(0,1)}},
      },
      loghist/.style={/pgfplots/.cd,
        corpus panel, ymin=0, ylabel={\% of kernels},
      },
    }
    \begin{axis}[corpus panel, at={(0,0)}, anchor=south west, width=2.8cm,
        title={(a) distinct kernels per program},
        ybar, bar width=5pt, xmin=-0.6, xmax=6.6, ymin=0, ymax=6000,
        ytick={0,2000,4000,6000}, yticklabels={0,2k,4k,6k},
        xtick=data, xticklabels from table={corpus-kernels.dat}{label},
        ylabel={programs},
        xmajorgrids=false,
        nodes near coords={\pgfmathprintnumber[fixed, precision=0, 1000 sep={,}, assume math mode=true]{\pgfplotspointmeta}},
        nodes near coords style={count label, anchor=south}]
      \addplot[fill=cA, draw=none] table[x=k, y=n] {corpus-kernels.dat};
    \end{axis}
    \begin{axis}[corpus panel, at={(4.2cm,0)}, anchor=south west, width=2.45cm,
        title={(b) source lines per program},
        xmin=0, xmax=300, ymin=0, ymax=30, ytick={0,10,20,30},
        xtick={0,100,200,300}, ylabel={\% of programs}]
      \addplot[const plot mark left, fill=cA, draw=none]
        table[x=edge, y=pct] {corpus-sloc.dat} \closedcycle;
      \draw[densely dotted, cF] (axis cs:92,0) -- (axis cs:92,30);
      \node[corner label] at (rel axis cs:1,1) {median 92};
    \end{axis}
    \begin{axis}[corpus panel, ylabel={\% of kernels},
        at={(0,-2.7cm)}, anchor=south west, width=2.8cm,
        title={(c) thread bundles per kernel},
        xmin=0, xmax=36, ymin=0, ymax=60, ytick={0,20,40,60},
        xtick={1,11,21,31}]
      \addplot[const plot mark left, fill=cA, draw=none]
        table[x=edge, y=pct] {corpus-bundles.dat} \closedcycle;
      \node[count label, anchor=south] at (axis cs:34,1) {$>$32};
    \end{axis}
    \begin{axis}[corpus panel, at={(4.95cm,-2.7cm)}, anchor=south west,
        width=1.85cm,
        title={(d) programs using \ldots},
        xbar, bar width=5pt, xmin=0, xmax=3600, ymin=0.4, ymax=4.6,
        xtick={0,1000,2000,3000}, xticklabels={0,1k,2k,3k}, xlabel={programs},
        ytick=data, yticklabels from table={corpus-features.dat}{label},
        ymajorgrids=false, xmajorgrids,
        nodes near coords={\pgfmathprintnumber[fixed, precision=0, 1000 sep={,}, assume math mode=true]{\pgfplotspointmeta}},
        nodes near coords style={count label, anchor=west}]
      \addplot[fill=cA, draw=none] table[x=n, y=y] {corpus-features.dat};
    \end{axis}
  \end{tikzpicture}
  \caption{Statistics about the programs in our dataset.}
  \label{fig:corpus}
\end{figure}

%% file: tab-bug-categories.tex
\begin{table}[t]
  \centering
  \footnotesize
  \setlength{\tabcolsep}{2pt}
  \renewcommand{\arraystretch}{1.08}
  \caption{The 600 bugs \dirigo\ reports in the \archive by category.
    ``Programs'' counts dataset entries,
    ``Sanitizer'' entries on which \texttt{compute-sanitizer} reports an error/warning,
    and ``Tasks'' distinct KernelBench tasks.}
  \label{tab:bug-categories}
  \begin{tabular}{@{}>{\raggedright\arraybackslash}p{0.62\columnwidth}
                    r r r@{}}
    \toprule
    Defect & Programs & Sanitizer & Tasks \\
    \midrule
    \multicolumn{4}{@{}l}{\textit{Synchronization} (\EXECUTE)} \\
    \quad Cross-warp shared-memory race
      & 131 & 131 & 26 \\
    \quad Warp-synchronous programming
      & 80 & 80 & 27 \\
    \quad Barrier under divergent control flow
      & 9 & 9 & 1 \\
    \addlinespace[2pt]
    \multicolumn{4}{@{}l}{\textit{Warp shuffle} (\EXECUTE)} \\
    \quad Full mask under a predicated warp
      & 67 & 11 & 24 \\
    \quad Shuffle partner outside the block
      & 6 & 3 & 1 \\
    \addlinespace[2pt]
    \multicolumn{4}{@{}l}{\textit{Memory safety} (\EXECUTE)} \\
    \quad Global OOB: shape or layout mismatch
      & 56 & 56 & 6 \\
    \quad Global OOB: boundary off-by-one
      & 11 & 1 & 8 \\
    \quad Uninitialized shared-memory read
      & 14 & 1 & 5 \\
    \addlinespace[2pt]
    \multicolumn{4}{@{}l}{\textit{Wrong computation} (\CHECK)} \\
    \quad Dropped, added, or reordered op
      & 135 & 0 & 3 \\
    \quad Wrong math function or constant
      & 12 & 0 & 4 \\
    \quad Wrong layout, thread mapping, or reduction
      & 79 & 3 & 6 \\
    \addlinespace[2pt]
    \midrule
    Total & 600 & 295 & 68 \\
    \bottomrule
  \end{tabular}
\end{table}

%% file: fig-eval-prevalence.tex
\begin{figure}[t]
  \centering
  \begin{tikzpicture}[count label/.style={font=\footnotesize, inner sep=1pt}]
    \begin{axis}[
      width=\columnwidth, height=0.48\columnwidth,
      ybar, bar width=0.34,
      xmin=-0.3, xmax=12.3,
      ymode=log, ymin=0.5, ymax=1500, log origin=infty,
      xtick={0,1,2,3,4,5,6,7,8,9,10,11,12},
      xticklabels={0,.01,.1,.2,.3,.4,.5,.6,.7,.8,.9,.99,1},
      ytick={1,10,100}, yticklabels={1,10,100},
      xlabel={fraction of output locations that reveal the \CHECK bug},
      ylabel={buggy programs},
      label style={font=\footnotesize},
      tick label style={font=\footnotesize},
      xtick pos=left, ytick pos=left,
      xtick align=outside, ytick align=outside,
      xtick style={draw=cF!60}, ytick style={draw=cF!60},
      axis line style={cF!60},
      ymajorgrids, grid style={cF!20},
      clip=false,
      unbounded coords=discard,
      legend style={at={(0.08,0.95)}, anchor=north west, font=\footnotesize,
                    draw=cF!40, fill=white, inner sep=1.5pt, row sep=-1pt,
                    legend cell align=left},
      legend image post style={scale=0.6},
    ]
      \addplot[fill=cA, draw=none, bar shift=-0.19]
        table[x=xc, y=total] {eval-prevalence-hist.dat};
      \addplot[fill=cB, draw=none, bar shift=0.19]
        table[x=xc, y=seq] {eval-prevalence-hist.dat};
      \legend{heuristic order, sequential order}
      \addplot[only marks, mark=none, bar shift=0, nodes near coords, point meta=explicit,
               nodes near coords style={count label, text=cA, anchor=south east}]
        table[x expr=\thisrow{xlab}-0.02, y=total, meta=total] {eval-prevalence-hist.dat};
      \addplot[only marks, mark=none, bar shift=0, nodes near coords, point meta=explicit,
               nodes near coords style={count label, text=cB, anchor=south west}]
        table[x expr=\thisrow{xlab}+0.02, y expr={max(\thisrow{seq},0.5)}, meta=seq] {eval-prevalence-hist.dat};
    \end{axis}
  \end{tikzpicture}
  \caption{
    The 226 programs with a bug found by \CHECK organized by how prevalent the bug is with sequential versus our heuristic search.}
  \label{fig:eval-prevalence}
\end{figure}
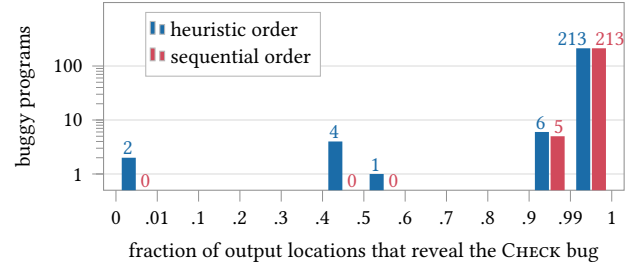

%% file: fig-eval-time.tex
\begin{figure}[t]
  \centering
  \begin{tikzpicture}[
      plateau label/.style={font=\footnotesize, anchor=south east,
                            inner sep=0pt, yshift=2pt},
      plateau label below/.style={plateau label, anchor=north east, yshift=-2pt},
    ]
    \begin{axis}[
      width=\columnwidth, height=0.46\columnwidth,
      xmin=0, xmax=241,
      ymin=0, ymax=660,
      xtick={0,60,120,180,240},
      ytick={0,200,400,600},
      xlabel={time to bug (s)},
      ylabel={bugs found},
      label style={font=\footnotesize},
      tick label style={font=\footnotesize},
      ymajorgrids, grid style={cF!20},
    ]
      \addplot[cA, thick, const plot]
        table[x=t, y=n_all] {eval-time.dat};
      \addplot[cB, thick, dashed, const plot]
        table[x=t, y=n_execute] {eval-time.dat};
      \addplot[cD, thick, densely dotted, const plot]
        table[x=t, y=n_check] {eval-time.dat};
      \node[plateau label below, text=cA] at (axis cs:239,600) {all bugs (600)};
      \node[plateau label, text=cB] at (axis cs:239,374) {\EXECUTE (374)};
      \node[plateau label, text=cD] at (axis cs:239,226) {\CHECK (226)};
    \end{axis}
  \end{tikzpicture}
  \caption{The number of bugs found versus wall-clock time
    thresholds, counted from the start of \TRACE (compilation excluded).
    }
  \label{fig:eval-time}
\end{figure}
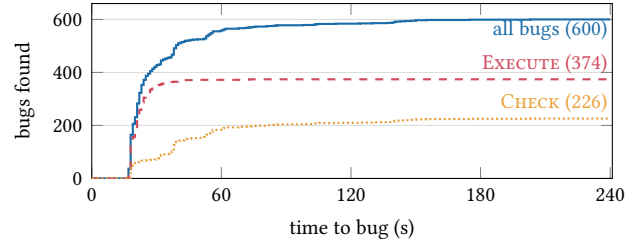

%% file: fig-eval-fullverif.tex
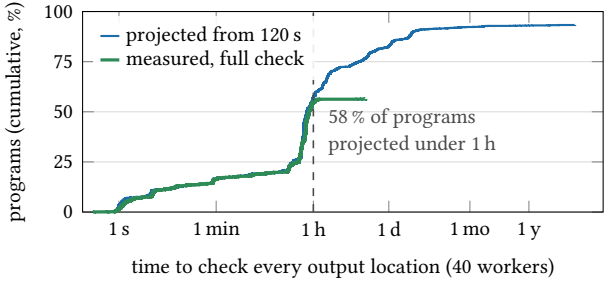
\begin{figure}[t]
  \centering
  \begin{tikzpicture}
    \begin{axis}[
      width=\columnwidth, height=0.50\columnwidth,
      xmode=log, log basis x=10,
      xmin=6e-5, xmax=2e5,
      ymin=0, ymax=100,
      xtick={0.000277778, 0.0166667, 1, 24, 730.5, 8766},
      xticklabels={1\,s, 1\,min, 1\,h, 1\,d, 1\,mo, 1\,y},
      xminorticks=false,
      ytick={0,25,50,75,100},
      xlabel={time to check every output location (40 workers)},
      ylabel={programs (cumulative, \%)},
      label style={font=\footnotesize},
      tick label style={font=\footnotesize},
      ymajorgrids, grid style={cF!20},
      legend style={font=\footnotesize, at={(0.02,0.98)}, anchor=north west,
                    draw=none, fill=white, fill opacity=0.9, text opacity=1,
                    row sep=-2pt, inner sep=2pt},
      legend image code/.code={\draw[#1] (0cm,0cm) -- (0.25cm,0cm);},
      legend cell align=left,
      clip=false,
    ]
      \addplot[cA, thick, const plot]
        table[x=hours, y=pct] {eval-fullverif-est.dat};
      \addlegendentry{projected from 120\,s}
      \addplot[cC, very thick, const plot]
        table[x=hours, y=pct] {eval-fullverif-actual.dat};
      \addlegendentry{measured, full check}
      \draw[dashed, semithick, cF] (axis cs:1,0) -- (axis cs:1,100)
        node[pos=0.40, anchor=west, align=left, font=\footnotesize, text=cF, xshift=2pt]
        {58\,\% of programs\\ projected under 1\,h};
    \end{axis}
  \end{tikzpicture}
  \caption{
    The share of programs with no bugs found for which \dirigo can check every output location within a given wall-clock time.}
  \label{fig:eval-fullverif}
\end{figure}

%% file: main.bbl
\begin{thebibliography}{32}

\ifx \showCODEN    \undefined \def \showCODEN     #1{\unskip}     \fi
\ifx \showISBNx    \undefined \def \showISBNx     #1{\unskip}     \fi
\ifx \showISBNxiii \undefined \def \showISBNxiii  #1{\unskip}     \fi
\ifx \showISSN     \undefined \def \showISSN      #1{\unskip}     \fi
\ifx \showLCCN     \undefined \def \showLCCN      #1{\unskip}     \fi
\ifx \shownote     \undefined \def \shownote      #1{#1}          \fi
\ifx \showarticletitle \undefined \def \showarticletitle #1{#1}   \fi
\ifx \showURL      \undefined \def \showURL       {\relax}        \fi
\providecommand\bibfield[2]{#2}
\providecommand\bibinfo[2]{#2}
\providecommand\natexlab[1]{#1}
\providecommand\showeprint[2][]{arXiv:#2}

\bibitem[Alur et~al\mbox{.}(2017)]%
        {alur2017gpudrano}
\bibfield{author}{\bibinfo{person}{Rajeev Alur}, \bibinfo{person}{Joseph
  Devietti}, \bibinfo{person}{Omar S.~Navarro Leija}, {and}
  \bibinfo{person}{Nimit Singhania}.} \bibinfo{year}{2017}\natexlab{}.
\newblock \showarticletitle{{GPUDrano}: Detecting Uncoalesced Accesses in {GPU}
  Programs}. In \bibinfo{booktitle}{\emph{Computer Aided Verification (CAV)}}
  \emph{(\bibinfo{series}{Lecture Notes in Computer Science})}.
  \bibinfo{publisher}{Springer International Publishing},
  \bibinfo{address}{Cham}, \bibinfo{pages}{507--525}.
\newblock
\href{https://doi.org/10.1007/978-3-319-63387-9_25}{doi:\nolinkurl{10.1007/978-3-319-63387-9_25}}

\bibitem[Alur et~al\mbox{.}(2022)]%
        {alur2022uncoalesced}
\bibfield{author}{\bibinfo{person}{Rajeev Alur}, \bibinfo{person}{Joseph
  Devietti}, \bibinfo{person}{Omar S.~Navarro Leija}, {and}
  \bibinfo{person}{Nimit Singhania}.} \bibinfo{year}{2022}\natexlab{}.
\newblock \showarticletitle{Static Detection of Uncoalesced Accesses in {GPU}
  Programs}.
\newblock \bibinfo{journal}{\emph{Formal Methods in System Design}}
  \bibinfo{volume}{60}, \bibinfo{number}{1} (\bibinfo{year}{2022}),
  \bibinfo{pages}{1--32}.
\newblock
\href{https://doi.org/10.1007/s10703-021-00362-8}{doi:\nolinkurl{10.1007/s10703-021-00362-8}}

\bibitem[Alur et~al\mbox{.}(2018)]%
        {alur2018blocksize}
\bibfield{author}{\bibinfo{person}{Rajeev Alur}, \bibinfo{person}{Joseph
  Devietti}, {and} \bibinfo{person}{Nimit Singhania}.}
  \bibinfo{year}{2018}\natexlab{}.
\newblock \showarticletitle{Block-Size Independence for {GPU} Programs}. In
  \bibinfo{booktitle}{\emph{Static Analysis Symposium (SAS)}}
  \emph{(\bibinfo{series}{Lecture Notes in Computer Science})}.
  \bibinfo{publisher}{Springer International Publishing},
  \bibinfo{address}{Cham}, \bibinfo{pages}{107--126}.
\newblock
\href{https://doi.org/10.1007/978-3-319-99725-4_9}{doi:\nolinkurl{10.1007/978-3-319-99725-4_9}}

\bibitem[Ansel et~al\mbox{.}(2024)]%
        {pytorch2compile2024}
\bibfield{author}{\bibinfo{person}{Jason Ansel}, \bibinfo{person}{Edward Yang},
  \bibinfo{person}{Horace He}, \bibinfo{person}{Natalia Gimelshein},
  \bibinfo{person}{Animesh Jain}, \bibinfo{person}{Michael Voznesensky},
  \bibinfo{person}{Bin Bao}, \bibinfo{person}{Peter Bell},
  \bibinfo{person}{David Berard}, \bibinfo{person}{Evgeni Burovski},
  \bibinfo{person}{Geeta Chauhan}, \bibinfo{person}{Anjali Chourdia},
  \bibinfo{person}{Will Constable}, \bibinfo{person}{Alban Desmaison},
  \bibinfo{person}{Zachary DeVito}, \bibinfo{person}{Elias Ellison},
  \bibinfo{person}{Will Feng}, \bibinfo{person}{Jiong Gong},
  \bibinfo{person}{Michael Gschwind}, \bibinfo{person}{Brian Hirsh},
  \bibinfo{person}{Sherlock Huang}, \bibinfo{person}{Kshiteej Kalambarkar},
  \bibinfo{person}{Laurent Kirsch}, \bibinfo{person}{Michael Lazos},
  \bibinfo{person}{Mario Lezcano}, \bibinfo{person}{Yanbo Liang},
  \bibinfo{person}{Jason Liang}, \bibinfo{person}{Yinghai Lu},
  \bibinfo{person}{C.~K. Luk}, \bibinfo{person}{Bert Maher},
  \bibinfo{person}{Yunjie Pan}, \bibinfo{person}{Christian Puhrsch},
  \bibinfo{person}{Matthias Reso}, \bibinfo{person}{Mark Saroufim},
  \bibinfo{person}{Marcos~Yukio Siraichi}, \bibinfo{person}{Helen Suk},
  \bibinfo{person}{Shunting Zhang}, \bibinfo{person}{Michael Suo},
  \bibinfo{person}{Phil Tillet}, \bibinfo{person}{Xu Zhao},
  \bibinfo{person}{Eikan Wang}, \bibinfo{person}{Keren Zhou},
  \bibinfo{person}{Richard Zou}, \bibinfo{person}{Xiaodong Wang},
  \bibinfo{person}{Ajit Mathews}, \bibinfo{person}{William Wen},
  \bibinfo{person}{Gregory Chanan}, \bibinfo{person}{Peng Wu}, {and}
  \bibinfo{person}{Soumith Chintala}.} \bibinfo{year}{2024}\natexlab{}.
\newblock \showarticletitle{{PyTorch} 2: Faster Machine Learning Through
  Dynamic {Python} Bytecode Transformation and Graph Compilation}. In
  \bibinfo{booktitle}{\emph{Proceedings of the 29th ACM International
  Conference on Architectural Support for Programming Languages and Operating
  Systems, Volume 2}} (La Jolla, CA, USA) \emph{(\bibinfo{series}{ASPLOS
  '24})}. \bibinfo{publisher}{Association for Computing Machinery},
  \bibinfo{address}{New York, NY, USA}, \bibinfo{pages}{929--947}.
\newblock
\showISBNx{9798400703850}
\href{https://doi.org/10.1145/3620665.3640366}{doi:\nolinkurl{10.1145/3620665.3640366}}

\bibitem[Baronio et~al\mbox{.}(2025)]%
        {kevin2025}
\bibfield{author}{\bibinfo{person}{Carlo Baronio}, \bibinfo{person}{Pietro
  Marsella}, \bibinfo{person}{Ben Pan}, \bibinfo{person}{Simon Guo}, {and}
  \bibinfo{person}{Silas Alberti}.} \bibinfo{year}{2025}\natexlab{}.
\newblock \bibinfo{title}{Kevin: Multi-Turn {RL} for Generating {CUDA}
  Kernels}.
\newblock
\showeprint[arxiv]{2507.11948}~[cs.LG]

\bibitem[Betts et~al\mbox{.}(2012)]%
        {gpuverify2012}
\bibfield{author}{\bibinfo{person}{Adam Betts}, \bibinfo{person}{Nathan Chong},
  \bibinfo{person}{Alastair~F. Donaldson}, \bibinfo{person}{Shaz Qadeer}, {and}
  \bibinfo{person}{Paul Thomson}.} \bibinfo{year}{2012}\natexlab{}.
\newblock \showarticletitle{{GPUVerify}: A Verifier for {GPU} Kernels}. In
  \bibinfo{booktitle}{\emph{Proceedings of the ACM International Conference on
  Object Oriented Programming Systems Languages and Applications}} (Tucson,
  Arizona, USA). \bibinfo{publisher}{Association for Computing Machinery},
  \bibinfo{address}{New York, NY, USA}, \bibinfo{pages}{113--132}.
\newblock
\showISBNx{9781450315616}
\href{https://doi.org/10.1145/2384616.2384625}{doi:\nolinkurl{10.1145/2384616.2384625}}

\bibitem[Bradbury et~al\mbox{.}(2018)]%
        {jax2018}
\bibfield{author}{\bibinfo{person}{James Bradbury}, \bibinfo{person}{Roy
  Frostig}, \bibinfo{person}{Peter Hawkins}, \bibinfo{person}{Matthew~James
  Johnson}, \bibinfo{person}{Chris Leary}, \bibinfo{person}{Dougal Maclaurin},
  \bibinfo{person}{George Necula}, \bibinfo{person}{Adam Paszke},
  \bibinfo{person}{Jake VanderPlas}, \bibinfo{person}{Skye Wanderman-Milne},
  {and} \bibinfo{person}{Qiao Zhang}.} \bibinfo{year}{2018}\natexlab{}.
\newblock \bibinfo{title}{{JAX}: composable transformations of {Python}+{NumPy}
  programs}.
\newblock \bibinfo{howpublished}{\url{http://github.com/jax-ml/jax}}.
\newblock

\bibitem[Chatterjee et~al\mbox{.}(2025)]%
        {chatterjee2026proofwright}
\bibfield{author}{\bibinfo{person}{Bodhisatwa Chatterjee},
  \bibinfo{person}{Drew Zagieboylo}, \bibinfo{person}{Sana Damani},
  \bibinfo{person}{Siva Hari}, {and} \bibinfo{person}{Christos Kozyrakis}.}
  \bibinfo{year}{2025}\natexlab{}.
\newblock \bibinfo{title}{ProofWright: Towards Agentic Formal Verification of
  CUDA}.
\newblock
\showeprint[arxiv]{2511.12294}~[cs.SE]
\urldef\tempurl%
\url{https://arxiv.org/abs/2511.12294}
\showURL{%
\tempurl}

\bibitem[Cogumbreiro et~al\mbox{.}(2021)]%
        {faial2021cav}
\bibfield{author}{\bibinfo{person}{Tiago Cogumbreiro}, \bibinfo{person}{Julien
  Lange}, \bibinfo{person}{Dennis Liew}, {and} \bibinfo{person}{Hannah
  Zicarelli}.} \bibinfo{year}{2021}\natexlab{}.
\newblock \showarticletitle{Checking Data-Race Freedom of GPU Kernels,
  Compositionally}. In \bibinfo{booktitle}{\emph{Computer Aided Verification}}
  \emph{(\bibinfo{series}{Lecture Notes in Computer Science})},
  \bibfield{editor}{\bibinfo{person}{Alexandra Silva} {and}
  \bibinfo{person}{K.~Rustan~M. Leino}} (Eds.). \bibinfo{publisher}{Springer
  International Publishing}, \bibinfo{address}{Cham},
  \bibinfo{pages}{403--426}.
\newblock
\showISBNx{978-3-030-81685-8}
\href{https://doi.org/10.1007/978-3-030-81685-8_19}{doi:\nolinkurl{10.1007/978-3-030-81685-8_19}}

\bibitem[Cogumbreiro et~al\mbox{.}(2024)]%
        {cogumbreiro2024fmsd}
\bibfield{author}{\bibinfo{person}{Tiago Cogumbreiro}, \bibinfo{person}{Julien
  Lange}, \bibinfo{person}{Dennis Liew}, {and} \bibinfo{person}{Hannah
  Zicarelli}.} \bibinfo{year}{2024}\natexlab{}.
\newblock \showarticletitle{Memory access protocols: certified data-race
  freedom for {GPU} kernels}.
\newblock \bibinfo{journal}{\emph{Formal Methods in System Design}}
  \bibinfo{volume}{63}, \bibinfo{number}{1--3} (\bibinfo{year}{2024}),
  \bibinfo{pages}{134--171}.
\newblock
\showISSN{1572-8102}
\href{https://doi.org/10.1007/s10703-023-00415-0}{doi:\nolinkurl{10.1007/s10703-023-00415-0}}

\bibitem[Collingbourne et~al\mbox{.}(2014)]%
        {collingbourne2014kleecl}
\bibfield{author}{\bibinfo{person}{Peter Collingbourne},
  \bibinfo{person}{Cristian Cadar}, {and} \bibinfo{person}{Paul~H.J. Kelly}.}
  \bibinfo{year}{2014}\natexlab{}.
\newblock \showarticletitle{Symbolic Crosschecking of Data-Parallel
  Floating-Point Code}.
\newblock \bibinfo{journal}{\emph{IEEE Transactions on Software Engineering}}
  \bibinfo{volume}{40}, \bibinfo{number}{7} (\bibinfo{year}{2014}),
  \bibinfo{pages}{710--737}.
\newblock
\href{https://doi.org/10.1109/TSE.2013.2297120}{doi:\nolinkurl{10.1109/TSE.2013.2297120}}

\bibitem[Collingbourne et~al\mbox{.}(2012)]%
        {collingbourne2012kleecl2}
\bibfield{author}{\bibinfo{person}{Peter Collingbourne},
  \bibinfo{person}{Cristian Cadar}, {and} \bibinfo{person}{Paul H.~J. Kelly}.}
  \bibinfo{year}{2012}\natexlab{}.
\newblock \showarticletitle{Symbolic Testing of OpenCL Code}. In
  \bibinfo{booktitle}{\emph{Hardware and Software: Verification and Testing}}
  \emph{(\bibinfo{series}{Lecture Notes in Computer Science})},
  \bibfield{editor}{\bibinfo{person}{Kerstin Eder}, \bibinfo{person}{Jo{\~a}o
  Louren{\c{c}}o}, {and} \bibinfo{person}{Onn Shehory}} (Eds.).
  \bibinfo{publisher}{Springer Berlin Heidelberg}, \bibinfo{address}{Berlin,
  Heidelberg}, \bibinfo{pages}{203--218}.
\newblock
\showISBNx{978-3-642-34188-5}
\href{https://doi.org/10.1007/978-3-642-34188-5_18}{doi:\nolinkurl{10.1007/978-3-642-34188-5_18}}

\bibitem[de~Moura and Bj{\o}rner(2008)]%
        {z3_2008}
\bibfield{author}{\bibinfo{person}{Leonardo de Moura} {and}
  \bibinfo{person}{Nikolaj Bj{\o}rner}.} \bibinfo{year}{2008}\natexlab{}.
\newblock \showarticletitle{{Z3}: An Efficient {SMT} Solver}. In
  \bibinfo{booktitle}{\emph{Tools and Algorithms for the Construction and
  Analysis of Systems}} \emph{(\bibinfo{series}{Lecture Notes in Computer
  Science})}. \bibinfo{publisher}{Springer Berlin Heidelberg},
  \bibinfo{address}{Berlin, Heidelberg}, \bibinfo{pages}{337--340}.
\newblock
\showISBNx{9783540788003}
\href{https://doi.org/10.1007/978-3-540-78800-3_24}{doi:\nolinkurl{10.1007/978-3-540-78800-3_24}}

\bibitem[Driscoll et~al\mbox{.}(2025)]%
        {volta2025}
\bibfield{author}{\bibinfo{person}{Benjamin Driscoll}, \bibinfo{person}{Kshitij
  Dubey}, \bibinfo{person}{Anjiang Wei}, \bibinfo{person}{Neeraj Kayal},
  \bibinfo{person}{Rahul Sharma}, {and} \bibinfo{person}{Alex Aiken}.}
  \bibinfo{year}{2025}\natexlab{}.
\newblock \bibinfo{title}{Equivalence Checking of {ML} {GPU} Kernels}.
\newblock
\showeprint[arxiv]{2511.12638}~[cs.PL]
\newblock
\shownote{v3, August 2026}.

\bibitem[Farooqui et~al\mbox{.}(2014)]%
        {farooqui2014lynx-symex}
\bibfield{author}{\bibinfo{person}{Naila Farooqui}, \bibinfo{person}{Karsten
  Schwan}, {and} \bibinfo{person}{Sudhakar Yalamanchili}.}
  \bibinfo{year}{2014}\natexlab{}.
\newblock \showarticletitle{Efficient Instrumentation of GPGPU Applications
  Using Information Flow Analysis and Symbolic Execution}. In
  \bibinfo{booktitle}{\emph{Proceedings of Workshop on General Purpose
  Processing Using GPUs}} (Salt Lake City, UT, USA)
  \emph{(\bibinfo{series}{GPGPU-7})}. \bibinfo{publisher}{Association for
  Computing Machinery}, \bibinfo{address}{New York, NY, USA},
  \bibinfo{pages}{19–27}.
\newblock
\showISBNx{9781450327664}
\href{https://doi.org/10.1145/2588768.2576782}{doi:\nolinkurl{10.1145/2588768.2576782}}

\bibitem[Godefroid et~al\mbox{.}(2005)]%
        {godefroid2005dart}
\bibfield{author}{\bibinfo{person}{Patrice Godefroid}, \bibinfo{person}{Nils
  Klarlund}, {and} \bibinfo{person}{Koushik Sen}.}
  \bibinfo{year}{2005}\natexlab{}.
\newblock \showarticletitle{DART: directed automated random testing}. In
  \bibinfo{booktitle}{\emph{Proceedings of the 2005 ACM SIGPLAN Conference on
  Programming Language Design and Implementation}} (Chicago, IL, USA)
  \emph{(\bibinfo{series}{PLDI '05})}. \bibinfo{publisher}{Association for
  Computing Machinery}, \bibinfo{address}{New York, NY, USA},
  \bibinfo{pages}{213–223}.
\newblock
\showISBNx{1595930566}
\href{https://doi.org/10.1145/1065010.1065036}{doi:\nolinkurl{10.1145/1065010.1065036}}

\bibitem[Hendrycks and Gimpel(2016)]%
        {hendrycks2016gelu}
\bibfield{author}{\bibinfo{person}{Dan Hendrycks} {and} \bibinfo{person}{Kevin
  Gimpel}.} \bibinfo{year}{2016}\natexlab{}.
\newblock \bibinfo{title}{Gaussian Error Linear Units ({GELUs})}.
\newblock
\showeprint[arxiv]{1606.08415}~[cs.LG]

\bibitem[Hong et~al\mbox{.}(2025)]%
        {autocomp2025}
\bibfield{author}{\bibinfo{person}{Charles Hong}, \bibinfo{person}{Sahil
  Bhatia}, \bibinfo{person}{Alvin Cheung}, {and} \bibinfo{person}{Yakun~Sophia
  Shao}.} \bibinfo{year}{2025}\natexlab{}.
\newblock \bibinfo{title}{Autocomp: A Powerful and Portable Code Optimizer for
  Tensor Accelerators}.
\newblock
\showeprint[arxiv]{2505.18574}~[cs.PL]
\urldef\tempurl%
\url{https://arxiv.org/abs/2505.18574}
\showURL{%
\tempurl}

\bibitem[Lange et~al\mbox{.}(2025a)]%
        {sakana2025}
\bibfield{author}{\bibinfo{person}{Robert~Tjarko Lange},
  \bibinfo{person}{Aaditya Prasad}, \bibinfo{person}{Qi Sun},
  \bibinfo{person}{Maxence Faldor}, \bibinfo{person}{Yujin Tang}, {and}
  \bibinfo{person}{David Ha}.} \bibinfo{year}{2025}\natexlab{a}.
\newblock \bibinfo{title}{The {AI} {CUDA} Engineer: Agentic {CUDA} Kernel
  Discovery, Optimization and Composition}.
\newblock
  \bibinfo{howpublished}{\url{https://huggingface.co/datasets/SakanaAI/AI-CUDA-Engineer-Archive}}.
\newblock
\newblock
\shownote{Sakana AI technical report and kernel archive}.

\bibitem[Lange et~al\mbox{.}(2025b)]%
        {sakana_robust2025}
\bibfield{author}{\bibinfo{person}{Robert~Tjarko Lange}, \bibinfo{person}{Qi
  Sun}, \bibinfo{person}{Aaditya Prasad}, \bibinfo{person}{Maxence Faldor},
  \bibinfo{person}{Yujin Tang}, {and} \bibinfo{person}{David Ha}.}
  \bibinfo{year}{2025}\natexlab{b}.
\newblock \bibinfo{title}{Towards Robust Agentic {CUDA} Kernel Benchmarking,
  Verification, and Optimization}.
\newblock
\showeprint[arxiv]{2509.14279}~[cs.SE]

\bibitem[Li and Gopalakrishnan(2010)]%
        {li2010pug}
\bibfield{author}{\bibinfo{person}{Guodong Li} {and} \bibinfo{person}{Ganesh
  Gopalakrishnan}.} \bibinfo{year}{2010}\natexlab{}.
\newblock \showarticletitle{Scalable SMT-based verification of GPU kernel
  functions}. In \bibinfo{booktitle}{\emph{Proceedings of the Eighteenth ACM
  SIGSOFT International Symposium on Foundations of Software Engineering}}
  (Santa Fe, New Mexico, USA) \emph{(\bibinfo{series}{FSE '10})}.
  \bibinfo{publisher}{Association for Computing Machinery},
  \bibinfo{address}{New York, NY, USA}, \bibinfo{pages}{187–196}.
\newblock
\showISBNx{9781605587912}
\href{https://doi.org/10.1145/1882291.1882320}{doi:\nolinkurl{10.1145/1882291.1882320}}

\bibitem[Li et~al\mbox{.}(2012)]%
        {gklee2012}
\bibfield{author}{\bibinfo{person}{Guodong Li}, \bibinfo{person}{Peng Li},
  \bibinfo{person}{Geof Sawaya}, \bibinfo{person}{Ganesh Gopalakrishnan},
  \bibinfo{person}{Indradeep Ghosh}, {and} \bibinfo{person}{Sreeranga~P.
  Rajan}.} \bibinfo{year}{2012}\natexlab{}.
\newblock \showarticletitle{{GKLEE}: Concolic Verification and Test Generation
  for {GPUs}}. In \bibinfo{booktitle}{\emph{Proceedings of the 17th ACM SIGPLAN
  Symposium on Principles and Practice of Parallel Programming}} (New Orleans,
  Louisiana, USA) \emph{(\bibinfo{series}{PPoPP '12})}.
  \bibinfo{publisher}{Association for Computing Machinery},
  \bibinfo{address}{New York, NY, USA}, \bibinfo{pages}{215--224}.
\newblock
\showISBNx{9781450311601}
\href{https://doi.org/10.1145/2145816.2145844}{doi:\nolinkurl{10.1145/2145816.2145844}}

\bibitem[Li et~al\mbox{.}(2026)]%
        {cudal1_2025}
\bibfield{author}{\bibinfo{person}{Xiaoya Li}, \bibinfo{person}{Albert Wang},
  \bibinfo{person}{Guoyin Wang}, \bibinfo{person}{Jiwei Li}, {and}
  \bibinfo{person}{Chris Shum}.} \bibinfo{year}{2026}\natexlab{}.
\newblock \showarticletitle{{CUDA-L1}: Improving {CUDA} Optimization via
  Contrastive Reinforcement Learning}. In \bibinfo{booktitle}{\emph{Proceedings
  of the Fourteenth International Conference on Learning Representations}} (Rio
  de Janeiro, Brazil) \emph{(\bibinfo{series}{ICLR '26})}.
  \bibinfo{publisher}{OpenReview.net}.
\newblock
\showeprint[arxiv]{2507.14111}~[cs.AI]

\bibitem[Liao et~al\mbox{.}(2025)]%
        {kernelevolve2025}
\bibfield{author}{\bibinfo{person}{Gang Liao}, \bibinfo{person}{Hongsen Qin},
  \bibinfo{person}{Ying Wang}, \bibinfo{person}{Alicia Golden},
  \bibinfo{person}{Michael Kuchnik}, \bibinfo{person}{Yavuz Yetim},
  \bibinfo{person}{Jia~Jiunn Ang}, \bibinfo{person}{Chunli Fu},
  \bibinfo{person}{Yihan He}, \bibinfo{person}{Samuel Hsia},
  \bibinfo{person}{Zewei Jiang}, \bibinfo{person}{Dianshi Li},
  \bibinfo{person}{Uladzimir Pashkevich}, \bibinfo{person}{Varna Puvvada},
  \bibinfo{person}{Feng Shi}, \bibinfo{person}{Matt Steiner},
  \bibinfo{person}{Ruichao Xiao}, \bibinfo{person}{Liyuan Li},
  \bibinfo{person}{Nathan Yan}, \bibinfo{person}{Xiayu Yu},
  \bibinfo{person}{Zhou Fang}, \bibinfo{person}{Roman Levenstein},
  \bibinfo{person}{Kunming Ho}, \bibinfo{person}{Haishan Zhu},
  \bibinfo{person}{Alec Hammond}, \bibinfo{person}{Richard Li},
  \bibinfo{person}{Ajit Mathews}, \bibinfo{person}{Kaustubh Gondkar},
  \bibinfo{person}{Abdul Zainul-Abedin}, \bibinfo{person}{Ketan Singh},
  \bibinfo{person}{Hongtao Yu}, \bibinfo{person}{Wenyuan Chi},
  \bibinfo{person}{Barney Huang}, \bibinfo{person}{Sean Zhang},
  \bibinfo{person}{Noah Weller}, \bibinfo{person}{Zach Marine},
  \bibinfo{person}{Wyatt Cook}, \bibinfo{person}{Carole-Jean Wu}, {and}
  \bibinfo{person}{Gaoxiang Liu}.} \bibinfo{year}{2025}\natexlab{}.
\newblock \bibinfo{title}{{KernelEvolve}: Scaling Agentic Kernel Coding for
  Heterogeneous {AI} Accelerators at {Meta}}.
\newblock
\showeprint[arxiv]{2512.23236}~[cs.LG]

\bibitem[Liew et~al\mbox{.}(2024)]%
        {liew2024faialaa}
\bibfield{author}{\bibinfo{person}{Dennis Liew}, \bibinfo{person}{Tiago
  Cogumbreiro}, {and} \bibinfo{person}{Julien Lange}.}
  \bibinfo{year}{2024}\natexlab{}.
\newblock \showarticletitle{Sound and Partially-Complete Static Analysis of
  Data-Races in GPU Programs}.
\newblock \bibinfo{journal}{\emph{Proc. ACM Program. Lang.}}
  \bibinfo{volume}{8}, \bibinfo{number}{OOPSLA2}, Article
  \bibinfo{articleno}{357} (\bibinfo{date}{Oct.} \bibinfo{year}{2024}),
  \bibinfo{numpages}{28}~pages.
\newblock
\href{https://doi.org/10.1145/3689797}{doi:\nolinkurl{10.1145/3689797}}

\bibitem[Mart{\'i}nez et~al\mbox{.}(2026)]%
        {martinez2026kuiper}
\bibfield{author}{\bibinfo{person}{Guido Mart{\'i}nez},
  \bibinfo{person}{Bastian K{\"o}pcke}, \bibinfo{person}{Jon{\'a}{\v{s}}
  Fiala}, \bibinfo{person}{Gabriel Ebner}, \bibinfo{person}{Tahina
  Ramananandro}, \bibinfo{person}{Michel Steuwer}, \bibinfo{person}{Tyler
  Sorensen}, {and} \bibinfo{person}{Nikhil Swamy}.}
  \bibinfo{year}{2026}\natexlab{}.
\newblock \showarticletitle{{Kuiper}: Correct and Efficient {GPU} Programming
  with Dependent Types and Separation Logic}.
\newblock \bibinfo{journal}{\emph{Proc. ACM Program. Lang.}}
  \bibinfo{volume}{10}, \bibinfo{number}{PLDI} (\bibinfo{date}{June}
  \bibinfo{year}{2026}), \bibinfo{pages}{830--854}.
\newblock
\showISSN{2475-1421}
\href{https://doi.org/10.1145/3808280}{doi:\nolinkurl{10.1145/3808280}}

\bibitem[Mart{\'i}nez and Sorensen(2026)]%
        {martinez2026nextfrontier}
\bibfield{author}{\bibinfo{person}{Guido Mart{\'i}nez} {and}
  \bibinfo{person}{Tyler Sorensen}.} \bibinfo{year}{2026}\natexlab{}.
\newblock \showarticletitle{The Next Frontier for {AI}-Generated Kernels:
  Correctness}. In \bibinfo{booktitle}{\emph{Proceedings of the 2026 ACM
  SIGPLAN International Workshop on Principles of Agentic Engineering}}
  (Boulder, CO, USA) \emph{(\bibinfo{series}{PAgE '26})}.
  \bibinfo{publisher}{Association for Computing Machinery},
  \bibinfo{address}{New York, NY, USA}, \bibinfo{pages}{25--37}.
\newblock
\href{https://doi.org/10.1145/3819802.3820580}{doi:\nolinkurl{10.1145/3819802.3820580}}

\bibitem[{METR}(2025)]%
        {metr2025}
\bibfield{author}{\bibinfo{person}{{METR}}.} \bibinfo{year}{2025}\natexlab{}.
\newblock \bibinfo{title}{Measuring Automated Kernel Engineering}.
\newblock
  \bibinfo{howpublished}{\url{https://metr.org/blog/2025-02-14-measuring-automated-kernel-engineering/}}.
\newblock

\bibitem[Ouyang et~al\mbox{.}(2025)]%
        {kernelbench2025}
\bibfield{author}{\bibinfo{person}{Anne Ouyang}, \bibinfo{person}{Simon Guo},
  \bibinfo{person}{Simran Arora}, \bibinfo{person}{Alex~L. Zhang},
  \bibinfo{person}{William Hu}, \bibinfo{person}{Christopher R{\'e}}, {and}
  \bibinfo{person}{Azalia Mirhoseini}.} \bibinfo{year}{2025}\natexlab{}.
\newblock \bibinfo{title}{{KernelBench}: Can {LLMs} Write Efficient {GPU}
  Kernels?}
\newblock
\showeprint[arxiv]{2502.10517}~[cs.LG]

\bibitem[Pereira et~al\mbox{.}(2016)]%
        {pereira2016esbmcgpu}
\bibfield{author}{\bibinfo{person}{Phillipe Pereira}, \bibinfo{person}{Higo
  Albuquerque}, \bibinfo{person}{Hendrio Marques}, \bibinfo{person}{Isabela
  Silva}, \bibinfo{person}{Celso Carvalho}, \bibinfo{person}{Lucas Cordeiro},
  \bibinfo{person}{Vanessa Santos}, {and} \bibinfo{person}{Ricardo Ferreira}.}
  \bibinfo{year}{2016}\natexlab{}.
\newblock \showarticletitle{Verifying CUDA programs using SMT-based
  context-bounded model checking}. In \bibinfo{booktitle}{\emph{Proceedings of
  the 31st Annual ACM Symposium on Applied Computing}} (Pisa, Italy)
  \emph{(\bibinfo{series}{SAC '16})}. \bibinfo{publisher}{Association for
  Computing Machinery}, \bibinfo{address}{New York, NY, USA},
  \bibinfo{pages}{1648–1653}.
\newblock
\showISBNx{9781450337397}
\href{https://doi.org/10.1145/2851613.2851830}{doi:\nolinkurl{10.1145/2851613.2851830}}

\bibitem[Saba et~al\mbox{.}(2026)]%
        {cutegen2026}
\bibfield{author}{\bibinfo{person}{Tara Saba}, \bibinfo{person}{Zhiyang Chen},
  \bibinfo{person}{Jikai~Jason Li}, \bibinfo{person}{Anne Ouyang},
  \bibinfo{person}{Xujie Si}, {and} \bibinfo{person}{Fan Long}.}
  \bibinfo{year}{2026}\natexlab{}.
\newblock \bibinfo{title}{{CuTeGen}: An {LLM}-Based Agentic Framework for
  Generation and Optimization of High-Performance {GPU} Kernels using {CuTe}}.
\newblock
\showeprint[arxiv]{2604.01489}~[cs.LG]

\bibitem[Wu et~al\mbox{.}(2020)]%
        {wu2020simulee}
\bibfield{author}{\bibinfo{person}{Mingyuan Wu}, \bibinfo{person}{Yicheng
  Ouyang}, \bibinfo{person}{Husheng Zhou}, \bibinfo{person}{Lingming Zhang},
  \bibinfo{person}{Cong Liu}, {and} \bibinfo{person}{Yuqun Zhang}.}
  \bibinfo{year}{2020}\natexlab{}.
\newblock \showarticletitle{Simulee: detecting CUDA synchronization bugs via
  memory-access modeling}. In \bibinfo{booktitle}{\emph{Proceedings of the
  ACM/IEEE 42nd International Conference on Software Engineering}} (Seoul,
  South Korea) \emph{(\bibinfo{series}{ICSE '20})}.
  \bibinfo{publisher}{Association for Computing Machinery},
  \bibinfo{address}{New York, NY, USA}, \bibinfo{pages}{937–948}.
\newblock
\showISBNx{9781450371216}
\href{https://doi.org/10.1145/3377811.3380358}{doi:\nolinkurl{10.1145/3377811.3380358}}

\end{thebibliography}
